\documentclass[12pt,a4paper]{article}

\usepackage{fixltx2e}

\usepackage[text={450pt,650pt},headheight={14.5pt},centering]{geometry}

\usepackage{lmodern}

\usepackage{graphicx}
\usepackage{booktabs}
\usepackage{subfig}
\usepackage[font=small]{caption}
\usepackage{tikz}

\usetikzlibrary{decorations.pathmorphing,decorations.pathreplacing}

\usepackage{cite}
\usepackage[bulletsep]{collref}

\usepackage{amsmath,amssymb}
\usepackage{mathtools}
\usepackage{bbm}
\usepackage{slashed}
\usepackage{braket}

\usepackage{array}
\usepackage{arydshln}
\usepackage{xspace}

\usepackage{hyperref}

\usepackage{quotes}

\numberwithin{equation}{section}

\makeatletter
 \let\old@startsection=\@startsection
 \let\oldl@section=\l@section
 \renewcommand{\@startsection}[6]{\old@startsection{#1}{#2}{#3}{#4}{#5}{#6\mathversion{bold}}}
 \renewcommand{\l@section}[2]{\oldl@section{\mathversion{bold}#1}{#2}}
\makeatother

\DeclareMathOperator{\Str}{Str}

\def\Xint#1{\mathchoice
  {\XXint\displaystyle\textstyle{#1}}%
  {\XXint\textstyle\scriptstyle{#1}}%
  {\XXint\scriptstyle\scriptscriptstyle{#1}}%
  {\XXint\scriptscriptstyle\scriptscriptstyle{#1}}%
  \!\int}
\def\XXint#1#2#3{{\setbox0=\hbox{$#1{#2#3}{\int}$}
    \vcenter{\hbox{$#2#3$}}\kern-.5\wd0}}
\def\pint{\;\Xint-}

\newcommand{\AdS}{\textup{AdS}}
\newcommand{\CFT}{\mathrm{CFT}}
\newcommand{\Sphere}{\mathrm{S}}
\newcommand{\Torus}{\mathrm{T}}
\newcommand{\CP}{\mathrm{CP}}

\newcommand{\alg}[1]{#1}
\newcommand{\grp}[1]{\mathrm{#1}}

\newcommand{\algD}[1]{\alg{d}(2,1;#1)}
\newcommand{\grpD}[1]{\grp{D}(2,1;#1)}

\newcommand{\grpSL}{\grp{SL}}

\newcommand{\algSU}{\alg{su}}
\newcommand{\grpSU}{\grp{SU}}

\newcommand{\algU}{\alg{u}}

\newcommand{\algPSU}{\alg{psu}}
\newcommand{\grpPSU}{\grp{PSU}}

\newcommand{\Integers}{\mathbb{Z}}

\newcommand{\order}{\mathcal{O}}

\newcommand{\superN}{\mathcal{N}}

\newcommand{\ie}{\textit{i.e.}\xspace}

\begin{document}


\thispagestyle{empty}
\begin{flushright}\footnotesize\ttfamily
Imperial-TP-OOS-2014-02 \\
UUITP-05/14
\end{flushright}
\vspace{2em}

\begin{center}
\textbf{\Large\mathversion{bold} Finite-gap equations for strings on $\AdS_3 \times \Sphere^3 \times \Torus^4$ with mixed 3-form flux}

\vspace{3em}

\textrm{\large Andrei Babichenko${}^1$, Amit Dekel${}^2$ and Olof Ohlsson Sax${}^3$}

\vspace{4em}

\begingroup\itshape

1. Department of Mathematics, Weizmann Institute of Science, Rehovot, 76100, Israel
\vspace{0.7em}

2. Department of Physics and Astronomy, Uppsala University, Box~516, SE-751~20~Uppsala, Sweden
\vspace{0.7em}

3. The Blackett Laboratory, Imperial College, London SW7 2AZ, United Kingdom 
\endgroup

\vspace{2em}

\texttt{babichenkoandrei@gmail.com, amit.dekel@physics.uu.se, o.olsson-sax@imperial.ac.uk}

\end{center}


\vspace{3em}

\begin{abstract}\noindent
  We study superstrings on $\AdS_3 \times \Sphere^3 \times \Torus^4$ supported by a combination of Ramond--Ramond and Neveu-Schwarz--Neveu-Schwarz three form fluxes, 
  and construct a set of finite-gap equations that describe the classical string spectrum. 
  Using the recently proposed all-loop S-matrix we write down the all-loop Bethe ansatz equations for the massive sector. In the thermodynamic limit the Bethe ansatz reproduces the finite-gap equations. 
  As part of this derivation we propose expressions for the leading order dressing phases. These phases differ from the well-known Arutyunov--Frolov--Staudacher phase that appears in the pure Ramond--Ramond case.
  We also consider the one-loop quantization of the algebraic curve and determine the one-loop corrections to the dressing phases. 
  Finally we consider some classical string solutions including finite size giant magnons and circular strings.
\end{abstract}

\newpage


\section{Introduction}

Integrability has proved a powerful tool in the study of the $\AdS/\CFT$ correspondence~\cite{Beisert:2010jr}, and has greatly enhanced our understanding of $\superN=4$ super-Yang Mills theory and of the dual string theory in $\AdS_5 \times \Sphere^5$.
More recently progress has been made in understanding $\AdS_3/\CFT_2$ from an integrability perspective~\cite{Babichenko:2009dk}. In particular type IIB string theory on $\AdS_3 \times \Sphere^3 \times \Torus^4$ and on $\AdS_3 \times \Sphere^3 \times \Sphere^3 \times \Sphere^1$ has been studied. An interesting feature of these theories is that the backgrounds can be supported by a combination of Ramond--Ramond (RR) and Neveu-Schwarz--Neveu-Schwarz (NSNS) fluxes. This provides us with a family of string backgrounds, parametrized by the coefficients $\kappa$ and $\chi$ sitting in front of the respective flux terms.

In the pure RR case ($\kappa=1$, $\chi=0$) there has been a lot of progress in applying integrability methods to $\AdS_3/\CFT_2$~\cite{Babichenko:2009dk,OhlssonSax:2011ms,Borsato:2012ud,Borsato:2012ss,Borsato:2013qpa,Borsato:2013hoa,Borsato:2014exa,Borsato:2014hja}. 
The pure NSNS case ($\kappa=0$, $\chi=1$), on the other hand, is in principle solvable using CFT methods~\cite{Maldacena:2000hw,Maldacena:2000kv}. 
String theory with mixed fluxes is a useful tool for understanding the connections between these two limits, and provides a unique opportunity to study the non-perturbative S-duality using well-developed integrability frameworks such as the Thermodynamic Bethe Ansatz.

In this paper we study classical and semi-classical type IIB string theory on an  $\AdS_3 \times \Sphere^3 \times \Torus^4$ background with mixed three-form fluxes using integrability methods. At the level of classical string theory, integrability manifests itself through the existence of a Lax representation of the equations of motions~\cite{Bena:2003wd}. The basis of this construction is a description of the background in terms of a supercoset with a $\Integers_4$ automorphism~\cite{Berkovits:1999zq}. A sigma model action based on the coset was originally constructed for $\AdS_5 \times \Sphere^5$~\cite{Metsaev:1998it}, and has later been used for other backgrounds supported by RR flux, such as $\AdS_4 \times \CP^3$~\cite{Arutyunov:2008if,Stefanski:2008ik}, $\AdS_3 \times \Sphere^3 \times \Torus^4$ and $\AdS_3 \times \Sphere^3 \times \Sphere^3 \times \Sphere^1$~\cite{Babichenko:2009dk,Zarembo:2010sg}. In~\cite{Cagnazzo:2012se} this construction was generalised to the case of mixed RR and NSNS fluxes by the inclusion of a topological Wess-Zumino (WZ) term. Such a generalisation is only possible for certain backgrounds where the supercoset is constructed by taking the quotient of a product of two identical supergroups by its diagonal bosonic subgroup, called a permutation supercoset.
These backgrounds include the $\AdS_3$ backgrounds mentioned above.

Remarkably, the mixed flux supercoset action introduced in~\cite{Cagnazzo:2012se} remains integrable also outside of the pure RR limit, provided the coefficients of the Green-Schwarz (GS) and WZ terms satisfy the relation $\chi^2 + \kappa^2 = 1$.
However, in the construction of~\cite{Cagnazzo:2012se}, the WZ term break the $\Integers_4$ symmetry.
This $\Integers_4$ symmetry breaking obscures a generalisation of integrability techniques developed 
in the context of the AdS/CFT correspondence where the $\Integers_4$ symmetry plays an essential role.
One important example is the construction of an algebraic curve and a set of finite-gap equations which has proven a powerful tool~\cite{Kazakov:2004qf,Kazakov:2004nh,Beisert:2004ag,Beisert:2005bm,Gromov:2007cd,Gromov:2007ky,Gromov:2008ec,Zarembo:2010yz}.
In this paper we generalize the action of the  $\Integers_4$ automorphism in a way that allows us to apply standard techniques for deriving the finite-gap equations~\cite{Babichenko:2009dk,Zarembo:2010yz}.

To apply integrability beyond the semi-classical level requires the construction of asymptotic Bethe ansatz equations.
The tree-level S-matrix describing scattering of massive world-sheet excitations in uniform light-cone gauge was derived in~\cite{Hoare:2013pma}. Based on those results a conjecture for the all-loop S-matrix was made in~\cite{Hoare:2013ida}. This S-matrix is very similar to the S-matrix in the pure RR case, which has been studied both using perturbative string theory~\cite{Sundin:2012gc,Rughoonauth:2012qd,Abbott:2012dd,Sundin:2013ypa,Abbott:2013ixa,Engelund:2013fja,Bianchi:2013nra,Sundin:2014sfa} and using symmetry arguments~\cite{Borsato:2012ud,Borsato:2012ss,Borsato:2013qpa,Borsato:2013hoa,Borsato:2014exa,Borsato:2014hja}. We will use the proposed all-loop S-matrix to construct a set of all-loop Bethe ansatz equations, which in the thermodynamic limit can be used to rederive the finite-gap equations.

In this paper we consider only the massive sector of the $\AdS_3 \times \Sphere^3 \times \Torus^4$ theory.
Recently, there has been progress in taking into account the massless modes of the theory both in the algebraic curve description~\cite{Lloyd:2013wza} and in the study of the all loop world-sheet S-matrix~\cite{Borsato:2014exa,Borsato:2014hja}.
In both case the analysis was made in the pure RR case.

The plan of the paper is as follows.
In section~\ref{sec:super-coset-B-field-integrability} we rewrite the action and Lax connection of~\cite{Cagnazzo:2012se} using a notation that allows us to employ the $\Integers_4$ automorphism to construct a set of finite-gap equations through the algebraic method of~\cite{Babichenko:2009dk,Zarembo:2010yz}. In section~\ref{sec:finite-gap-from-BA} we construct a set of Bethe ansatz equations from the conjectured all-loop S-matrix of~\cite{Hoare:2013ida}. By taking a thermodynamic limit of the Bethe equations at strong coupling we reproduce the finite-gap equations found in the previous section.
In section~\ref{sec:semi-classical} we discuss the semi-classical quantization of the algebraic curve and derive the spectrum of fluctuations around the ground state. We further make a prediction for the one-loop dressing phases appearing in the S-matrix. In section~\ref{sec:circular-string} we derive the quasi-momenta for a classic circular string in the mixed flux case. Finally, in section~\ref{sec:giant-magnon} we construct giant magnon solutions and calculate the leading finite-size correction to their classical dispersion relation. Some further technical details are given in the appendix.

\section{Integrability of supercosets with a B-field}
\label{sec:super-coset-B-field-integrability}

In this section we review the construction of the permutation supercosets Green-Schwarz sigma model action, carrying both RR and NSNS fluxes, introduced in~\cite{Cagnazzo:2012se}.
We are going to write the action in a slightly different notation than the one used in~\cite{Cagnazzo:2012se}, such that the $\Integers_4$ symmetry appears to be manifest.
We find this form to be more convenient when we later study the spectral problem.

Permutation supercosets are a special class of $\Integers_4$ supercosets where the supergroup is a direct sum of two identical supergroups, and the quotient is taken with respect to the diagonal of the bosonic subgroup, namely $\left(G\times G\right) / G_{\bar 0}^{\text{diag}}$.
A Green-Schwarz sigma model with only RR flux was constructed for such backgrounds in~\cite{Babichenko:2009dk}, e.g. for
\begin{equation}
\AdS_3 \times \Sphere^3 \simeq \frac{\grpPSU(1,1|2)^2}{\grpSU(1,1)\times \grpSU(2)},\quad
\AdS_3 \times \Sphere^3 \times \Sphere^3 \simeq \frac{\grpD{\alpha}^2}{\grpSU(1,1)\times \grpSU(2)\times \grpSU(2)},
\end{equation}
see also~\cite{Zarembo:2010sg}.
It is well known that for any $\Integers_4$ supercoset one can construct a Green-Schwarz action~\cite{Metsaev:1998it,Berkovits:1999zq} which is integrable~\cite{Bena:2003wd,Adam:2007ws,Beisert:2010kp,Magro:2010jx}.
Interestingly, for permutation supercosets one can further add a WZ piece which carries NSNS flux, which leave the model integrable, given that the relation between the coupling of WZ pieces carrying RR and NSNS fluxes ($\kappa$ and $\chi$ respectively) satisfies $\kappa^2+\chi^2 = 1$.

Before presenting the action, we introduce some notation.
Lets us denote an element $x \in \mathfrak{g}_L\oplus \mathfrak{g}_R$ of the superalgebra using the matrix natation
\begin{equation}
x = \left(
      \begin{array}{c||c}
        x_L & 0 \\  \hline\hline
        0 & x_R \\
      \end{array}
    \right)
\end{equation}
with $x_{L/R} \in \mathfrak{g}_{L/R}$, thus a double line separates the two superalgebras $\mathfrak{g}_{L/R}$.
As discussed above, for permutation supercosets $\mathfrak{g}_L = \mathfrak{g}_R$.
The $\Integers_4$ automorphism acts on $\mathfrak{g} = \mathfrak{g}_L\oplus \mathfrak{g}_R$ as~\cite{Babichenko:2009dk}
\begin{equation}
\Omega_4 (x) = \left(
      \begin{array}{c||c}
        x_R & 0 \\  \hline\hline
        0 & (-)^{F} x_L
      \end{array}
    \right),
\end{equation}
where $(-)^{F}$ equals $1$ or $-1$ if the sub-superalgebra element $x_L$ is even or odd respectively.
We could avoid using the $(-)^{F}$ symbol by using
\begin{equation}
\Omega_4 (x) = H x H^{-1},\quad
H = \left(
      \begin{array}{@{}c@{}||@{}c@{}}
        0 & \begin{array}{c|c}
         +1 & 0 \\ \hline
        0 & +1 \\
      \end{array} \\ \hline\hline
        \begin{array}{c|c}
        +1 & 0 \\ \hline
        0 & -1 \\
      \end{array} & 0 \\
      \end{array}
    \right),
\end{equation}
where one line separates between the even and odd blocks of the supermatrix.
Notice that by definition, we have $\Omega_4(A B) = \Omega_4(A) \Omega_4(B)$.
The Maurer-Cartan one-form is defined by $J= g^{-1} d g$, where $g\in G_L\times G_R$.
The Maurer-Cartan decomposes under the $\Integers_4$ automorphism  such that
\begin{equation}
  J = J_0 \oplus J_1 \oplus J_2 \oplus J_3.
\end{equation}
More explicitly, using the notation introduced above we have
\begin{equation}\label{eq:MConeformsgrading}
  \begin{aligned}
    J_0 & = \frac{1+(-)^F}{4}\left(
      \begin{array}{c||c}
        J_L+J_R & 0 \\ \hline\hline
        0 & J_L+J_R \\
      \end{array}
    \right) , \\
    J_1 & = \frac{1-(-)^F}{4}\left(
      \begin{array}{c||c}
        J_L-i J_R & 0 \\ \hline\hline
        0 & i(J_L-i J_R) \\
      \end{array}
    \right), \\
    J_2 & = \frac{1+(-)^F}{4}\left(
      \begin{array}{c||c}
        J_L-J_R & 0 \\ \hline\hline
        0 & -(J_L-J_R) \\
      \end{array}
    \right), \\
    J_3 & = \frac{1-(-)^F}{4}\left(
      \begin{array}{c||c}
        J_L+i J_R & 0 \\ \hline\hline
        0 & -i(J_L+ iJ_R) \\
      \end{array}
    \right),
  \end{aligned}
\end{equation}
where $J_{L,R} = g_{L,R}^{-1} d g_{L,R}$.

Before we write down the action, we introduce the matrix
\begin{equation}
  W = \left(
    \begin{array}{c||c}
      +1 & \phantom{+}0 \\ \hline\hline
      \phantom{+}0 & -1 \\
    \end{array}
  \right).
\end{equation}
This matrix acts on the superalgebra elements in an obvious way and has the properties
$W^2 = 1$ and $\Omega_4(W) = -W$ (namely, its grading is $2$) so acting with $W$ on an element of the superalgebra would change the grading by 2.
Furthermore, it commutes with all the superalgebra elements. Note that we allow the $\Integers_4$ generator to act on $W$ even though it is not a dynamic field.

Finally, we define the supertrace acting on an element $x\in \mathfrak{g}_L\oplus \mathfrak{g}_R$ as
\begin{equation}
  \Str x = \Str x_L + \Str x_R,
\end{equation}
where $\Str x_{L/R}$ is the usual supertrace definition for the superalgebra~\cite{Frappat:1996pb}.
This supertrace preserves the $\Integers_4$ automorphism structure, that is
\begin{equation}
  \Str x_{(i)} x_{(j)} = 0
\end{equation}
if $i+j \mod 4\neq 0$, where $i,j=0,..3$ are the grading indices.

Using the notation introduced above, we write the mixed flux action~\cite{Cagnazzo:2012se}
\begin{equation}\label{eq:action2}
  \begin{split}
    S=&\frac{1}{2}\int_{\mathcal{M}}\Str \left(J_2\wedge\ast J_2+\kappa J_1\wedge J_3\right) \\
    &+\chi\int_{\mathcal{B}}\Str W \left(\frac{2}{3}J_2\wedge J_2\wedge J_2+J_1\wedge J_3\wedge J_2+J_3\wedge J_1\wedge J_2\right).
  \end{split}
\end{equation}
By using the matrix $W$, we have written the action in a form where the $\Integers_4$ automorphism leaves it invariant. This is not a physical symmetry since the automorphism acts non-trivially on the non-dynamic matrix $W$, which is equivalent to sending the coupling $\chi$ to $-\chi$.\footnote{%
  We thank Konstantin Zarembo for discussions on this point.
} %
Throughout the paper we will use this definition of the $\Integers_4$ automorphism, including the action on $W$, as it will be useful for studying the spectral problem.
Moreover, we do not have to introduce a special definition of the supertrace for the second integral as in~\cite{Hoare:2013pma}.
The equations of motion are given by
\begin{equation}
  d \ast K + \ast K \wedge J + J\wedge \ast K = 0
\end{equation}
where $K \equiv K_{\text{GS}}+K_{\text{WZ}}$, and
\begin{equation}
  \begin{aligned}
    \ast K_{\text{GS}} &= - \left(\ast 2 J_2+\kappa\left(J_3-J_1\right)\right), \\
    \ast K_{\text{WZ}} &= -\chi W \left(2 J_2 + J_1 +J_3\right).
  \end{aligned}
\end{equation}
In this form it is easy to see that $k \equiv g K g^{-1}$ satisfies
\begin{equation}\label{eq:Noether}
  d \ast k = 0,
\end{equation}
so $k$ is the Noether current.

In what follows, we assume that $\kappa$ and $\chi$ are related by $\kappa^2+\chi^2 = 1$ which was shown to be the condition for the theory to be integrable~\cite{Cagnazzo:2012se}. Thus, in some cases it will be more convenient to use only one independent parameter. Throughout the paper we will use intensively the parameter
\begin{equation}
s = s(\chi) = \sqrt{\frac{1+\chi}{1-\chi}},
\end{equation}
for reasons which will become clear in the following sections.
Sometimes we also find it convenient to use the angle $\psi$ such that $\kappa = \cos\psi$ and $\chi = \sin\psi$.
The later convention is useful in putting the action in a more compact form, which also unifies the two WZ terms into one which takes a similar form to the well known pure RR WZ term~\cite{Metsaev:1998it}.
Let us introduce a new set of fermionic currents by rotating $J_1$ and $J_3$,
\begin{equation}
  \begin{pmatrix}
    Q_1 \\ Q_3
  \end{pmatrix}
  =
  \begin{pmatrix}
    \phantom{+}\cos\frac{\psi W}{2} & \sin\frac{\psi W}{2} \\
    -\sin\frac{\psi W}{2} & \cos\frac{\psi W}{2}
  \end{pmatrix}
  \begin{pmatrix}
    J_1 \\ J_3
  \end{pmatrix} .
\end{equation}
The presence of $W$ guarantees that the gradings of $Q_1$ and $Q_3$ matches the grading denoted by the subscripts.
Using these currents we have
\begin{multline}
  \int_\mathcal{B} \Str \bigg(J_2\wedge (Q_1\wedge Q_1 -Q_3\wedge Q_3)\bigg)
  \\
  =\frac{\kappa }{2}\int_\mathcal{M}\Str \left(J_1\wedge J_3\right)
  +\chi\int_\mathcal{B}\Str W \left(J_1\wedge J_3\wedge J_2+J_3\wedge J_1\wedge J_2\right),
\end{multline}
which is nothing but the ``fermionic'' part of the WZ terms.
In order to get this result one has to use the form of the GS WZ term before integrating the total derivative~\cite{Metsaev:1998it,Arutyunov:2009ga}.
In this form the relation to the pure RR case is manifest.
By further defining $H_0 =  \sin (W\psi) J_2$ we can rewrite the action as
\begin{equation}\label{eq:alternativeform}
  S=\frac{1}{2}\int_\mathcal{M}\Str \left(J_2\wedge\ast J_2\right)
  +
  \int_\mathcal{B} \Str \bigg(J_2\wedge \left(\frac{2}{3}H_0\wedge J_2 + Q_1\wedge Q_1 -Q_3\wedge Q_3\right)\bigg).
\end{equation}

\subsection{Flat connection}
\label{sec:flat-connection}

The flat connection associated with the action~\eqref{eq:action2} was found in~\cite{Cagnazzo:2012se}.
However, as we shall see below, the description in~\cite{Cagnazzo:2012se} is not complete.
In this section we repeat the derivation using our notations and using different parametrization for the spectral parameter than the one used in~\cite{Cagnazzo:2012se}. We shall also discuss the properties of the flat connection emphasizing the difference with the vanishing NSNS flux case.

We start with the usual ansatz for the flat connection~\cite{Bena:2003wd}
\begin{equation}
  A = J_0 + \gamma_2 J_2 + \gamma_\ast \ast J_2 + \gamma_1 J_1 + \gamma_3 J_3.
\end{equation}
Requiring $d A + A \wedge A = 0$ yields the following set of equations
\begin{equation}\label{eq:cons_eq}
  \begin{aligned}
    \gamma_1^2-\gamma_2+\gamma_\ast \kappa &= 0, \\
    -\gamma_2+\gamma_3^2-\gamma_\ast \kappa &= 0, \\
    -1+\gamma_1 \gamma_3+\gamma_\ast \chi W &= 0, \\
    -1+\gamma_2^2-\gamma_\ast^2+2 \gamma_\ast \chi W &= 0, \\
    -\gamma_1+\gamma_2 \gamma_3-\gamma_3 \gamma_\ast \kappa +\gamma_1 \gamma_\ast \chi W &= 0, \\
    \gamma_1 \gamma_2-\gamma_3+\gamma_1 \gamma_\ast \kappa +\gamma_3 \gamma_\ast \chi W &= 0,
  \end{aligned}
\end{equation}
where we have used the equations of motion and the Maurer-Cartan equations.
Because of the presence of the $W$ matrix in the equations,
in contrast to previous works we allow the $\gamma_i$'s to be matrices such that $\gamma_i = \alpha_i \mathbb{I}+ \beta_i W$, where $\mathbb{I}$ is the unit matrix (this is consistent with the set of equations in~\eqref{eq:cons_eq}).
In the upper-left block we get the same set of equations as in~\eqref{eq:cons_eq} with $W$ replaced by $1$ and $\gamma_i$ by $\alpha_i+\beta_i$.
In the lower-right block we get the same set of equations as in~\eqref{eq:cons_eq} with $W$ replaced by $-1$ and $\gamma_i$ by $\alpha_i-\beta_i$.
The solutions for the first set is~\cite{Cagnazzo:2012se}
\begin{equation}
  \begin{aligned}
    \delta_\ast(\chi) = \alpha_\ast + \beta_\ast & = \chi \pm \sqrt{\delta_2^2(\chi) - \kappa^2}, \\
    \delta_1(\chi) = \alpha_1 + \beta_1 & = \pm \sqrt{\delta_2(\chi) - \kappa \delta_\ast(\chi)}, \\
    \delta_3(\chi) = \alpha_3 + \beta_3 & = \pm \sqrt{\delta_2(\chi) + \kappa \delta_\ast(\chi)},
  \end{aligned}
\end{equation}
given that $\kappa^2 = 1 - \chi^2$. Similarly, we have a solution to $\alpha_i-\beta_i$ with $\chi\to -\chi$.
The signs in the above equation are not necessarily synchronized.
Thus, eventually the flat connection can be written as follows
\begin{equation}\label{eq:flatconnection}
  A(x) = \frac{1}{2}\left(
    (\delta_i(\chi)+\delta_i(-\chi))J_i
    +(\delta_i(\chi)-\delta_i(-\chi))W J_i
  \right)
\end{equation}
where the index $i$ runs over $0,1,2,3,\ast$, and $\delta_0(\chi) = 1$.
The flat connection is not unique, since the flatness equation is invariant under the gauge transformation
\begin{equation}
  A \to A' = h A h^{-1} - dh h^{-1},
\end{equation}
for any $h$.
We would like the flat connection to transform covariantly under gauge transformations, in order to be able to read the Noether current.
Choosing $h=g$ will eliminate the $J_0$ dependence
\begin{equation}
  A \to a = g (A - J) g^{-1}.
\end{equation}

Next, we would like to parameterize $\delta_2$ in terms of a spectral parameter $x$.
In principle, we can choose any function of $\delta_2(x)$ (which could also depend on $\chi$), however, we would like the flat connection to have several properties which we list below.
First, we would like the flat connection to coincide with the ``standard'' flat connection used for the algebraic curve analysis~\cite{Beisert:2005bm} when the NSNS flux is zero, \ie, $\delta_2(x) \to \frac{x^2 + 1}{x^2 - 1}$ as $\chi \to 0$.
We would also like to keep the properties
\begin{equation}
  a(x) \sim \frac{k}{x} + \dotsb,
\end{equation}
as $x \to \infty$ and where $k$ is the Noether current~\eqref{eq:Noether}, as well as the $\Integers_4$ symmetry relation
\begin{equation}\label{eq:flat_symm}
  A\left(\frac{1}{x}\right) = \Omega_4\left( A(x) \right).
\end{equation}
Finally, we require $\delta_2$ and $\delta_\ast$ to be rational functions of the spectral parameter.

The following parametrization gives the desired properties\footnote{This choice is different than the one given in~\cite{Cagnazzo:2012se}. Both choices reduce to the ``standard'' flat connection when $\chi\to 0$. However, expanding the flat connection of~\cite{Cagnazzo:2012se} at infinity does not yield the Noether current.}
\begin{equation}
  \begin{aligned}\label{eq:Laxcoefficients}
    \delta_2 &= \frac{\left(x^2+1\right) \kappa }{\left(x^2-1\right) \kappa -2 x \chi }, \quad &
    \delta_1 &= (x+1)\sqrt{\frac{\kappa }{\left(x^2-1\right) \kappa -2 x \chi }}, \\
    \delta_\ast &= -\frac{2 x \kappa }{(x \kappa -\chi )^2-1}, \quad &
    \delta_3 &=  (x-1)\sqrt{\frac{\kappa }{\left(x^2-1\right) \kappa -2 x \chi }}.
  \end{aligned}
\end{equation}
This parametrization is not unique given the above requirements, another requirement we used is that the poles coincide with the ones we get from the Bethe ansatz analysis below.
Notice that
\begin{equation}
  \delta_i\left(\frac{1}{x},-\chi\right)J_i = \delta_i(x,\chi)\Omega_4\left( J_i \right),\quad
  \Omega_4(W)=-W,
\end{equation}
so by using the representation~\eqref{eq:flatconnection} for the flat connection, equation~\eqref{eq:flat_symm} is obviously satisfied.
Comparing with~\eqref{eq:MConeformsgrading}, the flat connection can be written as
\begin{equation}
A(x) = \left(
  \begin{array}{c||c}
    \hat A(x) & 0 \\ \hline\hline
    0 & (-)^F\hat A(1/x) \\
  \end{array}
\right).
\end{equation}
One can check that indeed the $x\to\infty$ expansion gives the Noether current~\eqref{eq:Noether}
\begin{equation}\label{eq:largeXexp}
  a(x \to \infty) \simeq -\frac{1}{\kappa x}g\left(-2W \chi J_2+ 2\ast J_2-(\kappa+W \chi)J_1+(\kappa-W \chi)J_3\right)g^{-1} + \mathcal{O}(1/x^2).
\end{equation}
Notice that the Noether current is rescaled by $1/\kappa$. This can be cured by rescaling $x\to x/\kappa$, which also simplifies the coefficients of the flat connection\footnote{\label{fn:other-parametrization}
Using this parametrization the coefficient of the flat connection take the form
\begin{equation}
  \begin{aligned}
    \delta_2 &= \frac{1+x^2-\chi^2}{(x-\chi)^2 -1}, \quad &
    \delta_1 &= \frac{x+\kappa}{\sqrt{(x-\chi)^2 -1}}, \\
    \delta_\ast &= -\frac{2 x}{(x-\chi)^2 -1}, \quad &
    \delta_3 &=  \frac{x-\kappa}{\sqrt{(x-\chi)^2 -1}}.
  \end{aligned}
\end{equation}
}. However, the price is that the symmetry $x\to 1/x$ is then modified to $x\to \kappa^2/x$.

In the case where there is no NSNS flux the flat connection has poles at $x = \pm 1$.
In our case the poles are shifted such that $\hat A(x)$ defined above has poles at
\begin{equation}
  \hat s_\pm = \pm \sqrt{\frac{1\pm\chi}{1\mp\chi}}.
\end{equation}
Expanding around the poles $\hat s_\pm$ we get
\begin{align}
  A(x) \simeq&
  \frac{1}{2}\left(\mathbb{I}+W\right)\bigg(\frac{\hat s_{\pm }}{x-\hat s_{\pm }}(J_2\pm \ast J_2)+\sqrt{\frac{\hat s_{\pm } }{x-\hat s_{\pm }}}\left(\frac{ \left(1+\hat s_{\pm }\right)}{\sqrt{\left(1+\hat s_{\pm }{}^2\right)}}J_1+\frac{\left(-1+\hat s_{\pm }\right)}{\sqrt{\left(1+\hat s_{\pm }{}^2\right) }}J_3\right) \nonumber \\
  &+\left(J_0+\frac{ \hat s_{\pm }{}^2}{1+\hat s_{\pm }{}^2}J_2\mp \frac{1}{1+\hat s_{\pm }{}^2}\ast J_2\right) \nonumber \\
  &+\sqrt{\frac{x-\hat s_{\pm } }{\hat s_{\pm }}}\left(\frac{ \left(2-\hat s_{\pm }+\hat s_{\pm }{}^2\right) \hat s_{\pm }}{2 \left(1+\hat s_{\pm }{}^2\right){}^{3/2}}J_1+\frac{ \left(2+\hat s_{\pm }+\hat s_{\pm }{}^2\right) \hat s_{\pm }}{2 \left(1+\hat s_{\pm }{}^2\right){}^{3/2}}J_3\right) \nonumber \\
  &+\frac{x-\hat s_{\pm }}{\hat s_{\pm }}\left(\frac{ \hat s_{\pm } }{1+\hat s_{\pm }{}^2}\right){}^2(J_2\pm \ast J_2)+ \dotsb \bigg).
\end{align}
Because of the $\Integers_4$ symmetry we also have poles at $x=1/\hat s_{\pm}$. Defining $\check s_\pm = \mp\sqrt{\frac{1\pm\chi}{1\mp\chi}}$, and expanding around around $\check s_\pm$ we get the same expansion as above with
$W\to -W$,
$\hat s_\pm\to -\check s_\pm$
and $\ast J_2 \to -\ast J_2$.
It will turn out to be convenient to define the poles in term of a new variable $s$ instead of $\chi$ such that
\begin{equation}
  \hat s_+ = s,\quad
  \hat s_- = -1/s,\quad
  \check s_+ = -s,\quad
  \check s_- = 1/s.
\end{equation}
Then, the residues are given by
\begin{equation}
  \begin{aligned}\label{eq:residues}
    A\left(x\simeq s\right) & = \frac{1}{2}\left(\mathbb{I}+W\right)\frac{s}{x-s}(J_2 + \ast J_2) + \dotsb \\
    A\left(x\simeq -s^{-1}\right) & = \frac{1}{2}\left(\mathbb{I}+W\right)\frac{-s^{-1}}{x+s^{-1}}(J_2 - \ast J_2) + \dotsb \\
    A\left(x\simeq -s\right) & = \frac{1}{2}\left(\mathbb{I}-W\right)\frac{-s}{x+s}(J_2 - \ast J_2) + \dotsb \\
    A\left(x\simeq s^{-1}\right) & = \frac{1}{2}\left(\mathbb{I}-W\right)\frac{s^{-1}}{x-s^{-1}}(J_2 + \ast J_2)+ \dotsb
  \end{aligned}
\end{equation}

Let us also introduce the $z$ variable by $x = \frac{s+z^2}{1-s z^2}$, similar in spirit to the transformation given in~\cite{Beisert:2005bm}, which yields
\begin{equation}
  \frac{dx}{(x-s)(x+s^{-1})} = \frac{2}{s+s^{-1}}\frac{dz}{z}.
\end{equation}
Using these variables, the flat connection coefficients are given by
\begin{equation}
  \begin{aligned}
    \delta_2 &= \frac{\left(1+z^4\right) \cos\psi}{2 z^2}, \quad &
    \delta_1 &= \frac{\sqrt{\cos\psi} \left(\cos\frac{\psi}{2}-z^2 \sin\frac{\psi}{2}\right)}{z}, \\
    \delta_\ast &= \frac{\left(z^4-1\right) \cos\psi}{2 z^2}+\sin\psi, \quad &
    \delta_3 &=  \frac{\sqrt{\cos\psi} \left(z^2 \cos\frac{\psi}{2}+\sin\frac{\psi}{2}\right)}{z}.
  \end{aligned}
\end{equation}
These coefficients satisfy $i^k \delta_k(i z,-\psi) = \delta_k(z,\psi)$, and the poles at $s$ and $s^{-1}$ are mapped to $0$ and $\infty$ respectively.

As a final comment, as is the case in~\cite{Cagnazzo:2012se}, taking the pure NSNS limit by sending $\chi\to 1$ (and $\kappa\to 0$), the flat connection degenerates, \ie, the only nontrivial coefficient of the flat connection is $\delta_\ast = 1$.
However, taking the limit for a general choice of a parametrization of $\delta_2$ yields the equations
\begin{equation}
  W \delta_\ast = 1 -\delta_2,\quad
  \delta_1 = \delta_3 = \pm\sqrt{\delta_2},\quad \text{or}\quad
  W \delta_\ast = 1 +\delta_2,\quad
  \delta_1 = -\delta_3 = \pm\sqrt{\delta_2}.
\end{equation}
These equations do not generally yield a degenerate flat connection. An example of a parametrization that does not degenerate in the $\chi \to 1$ limit is given in footnote~\ref{fn:other-parametrization}.

\subsection{Finite-gap integration}
\label{sec:finite-gap-integration}

In this section we derive the finite-gap equations following~\cite{Babichenko:2009dk,Zarembo:2010yz}.
As will be shown below, the construction requires some modifications, mainly because of the poles shift.
Starting with the flat connection given in~\eqref{eq:flatconnection} and using the coefficients introduced in~\eqref{eq:Laxcoefficients}, we define the monodromy matrix
\begin{equation}
  M(x,\chi) = P \exp \int_0^{2\pi} A_\sigma(x,\chi) =  U^{-1}(x,\chi)\exp(p_i(x,\chi) H_i)U(x,\chi)
\end{equation}
where $H_l$ is a Cartan basis of the superalgebra.
$p_l(x,\chi)$ is the quasi-momentum, it is gauge invariant and defined up to Weyl group transformations and shifts of integers time $2\pi$.
By construction, the monodromy matrix is given by two blocks.
We shall denote the quasi-momenta in the upper-left block by $\hat p(x)$ and the quasi-momenta in the lower-right block by $\check p_l(x)$. Un-hatted/checked quasi-momenta will denote either hatted or checked quasi-momenta.

For large $x$ we get~\eqref{eq:largeXexp}
\begin{equation}
  A = g^{-1}\left(d + \frac{1}{\kappa x} \ast k\right) g+\mathcal{O}(1/x^2),
\end{equation}
where $k$ is the Noether current defined above, $d\ast k=0$.
This implies that
\begin{equation}
  p_l(x) = \frac{1}{\kappa x} q_l+\mathcal{O}(1/x^2)
\end{equation}
since the quasi-momenta are gauge invariant and where $q_l$ are the global Noether charges.

As discussed in the previous section, the Lax connection has four poles at $\hat s_+=s$, $\hat s_-=-1/s$, $\check s_+=-s$ and $\check s_-=1/s$ with $s= \sqrt{\frac{1+\chi}{1-\chi}}=\frac{1+\chi}{\kappa}=\frac{1+\sin\psi}{\cos\psi}$.
The monodromy matrix is a meromorphic function of $x$ with possible singularities located at these poles.
Strictly speaking $\hat p(x)$ will have poles at $\hat s_{\pm}$ and $\check p(x)$ at $\check s_{\mp}$.

The Weyl reflections of the superalgebra act on the quasi-momenta as
\begin{equation}
  p_l(x)\to p_l(x) - A_{lm} p_m(x),
\end{equation}
where $A$ is the Cartan matrix.
when encircling a branch point $a_{l,i}$ on the complex plane $x$, the quasi-momentum changes as
\begin{equation}
  p_l(x)\to p_l(x) - A_{lm} p_m(x)+2\pi n_{l,i}.
\end{equation}
For fermionic roots $A_{ll}=0$, so $p_l \to p_l + \dotsb$ which implies a logarithmic branch point.
For bosonic roots $A_{ll}=2$, so $p_l \to -p_l +\dotsb$ which implies a square root branch point.
Let us denote the bosonic square root cuts as $C_{l,i}$ so that
\begin{equation}\label{eq:disc}
  A_{lm}\slashed{p}_m = 2\pi n_{l,i},\quad x\in C_{l,i}.
\end{equation}
Close to the poles, the Lax-connection is given by~\eqref{eq:residues},
so the quasi-momenta is given by
\begin{equation}
  \begin{aligned}
    \hat p_l(x\simeq \hat s_\pm) &\to \pm \frac{\hat s_\pm}{2} \frac{\hat \kappa_l \pm 2\pi \hat m_l }{x-\hat s_{\pm}} + \dotsb , \\
    \check p_l(x\simeq \check s_\pm) &\to \mp \frac{\hat s_\pm}{2} \frac{\check \kappa_l \mp 2\pi \check m_l }{x+\hat s_{\pm}} + \dotsb ,
  \end{aligned}
\end{equation}
where we have used~\eqref{eq:MConeformsgrading} and~\eqref{eq:residues} to show that the residues are the same for the $\hat s_\pm$ and $\check s_\pm$ poles.

The Lax-connection has the $\Integers_4$ symmetry property~\eqref{eq:flat_symm}
\begin{equation}
  \Omega_4(A(x)) = A\left(1/x\right),
\end{equation}
so using $\Omega_4(A B) = \Omega_4(A) \Omega_4(B)$, we also have
\begin{equation}
  \Omega_4(M(x)) = M\left(1/x\right).
\end{equation}
Given
\begin{equation}
  \Omega_4(H_l) = H_m S_{ml}
\end{equation}
where the eigenvalues of $S_{ml}$ are $\pm 1$ since $\Omega^2 = (-)^F$, we should also have
\begin{equation}\label{eq:quasimomentasymm}
  p_l(1/x) = S_{lm} p_m(x).
\end{equation}
Finally, the Virasoro constraint are not modified by the WZ term, so the null condition is the same as in~\cite{Zarembo:2010yz}\footnote{Here we consider only the massive sector. In order to include the massless modes contribution, the condition~\eqref{eq:Virasoro} should be relaxed, allowing for the contribution from other fields which are not included in the coset~\cite{Zarembo:2010yz,Lloyd:2013wza}.}
\begin{equation}\label{eq:Virasoro}
  (\kappa_l\pm 2\pi m_l)A_{l k}(\kappa_k\pm 2\pi m_k) = 0.
\end{equation}
Next, we write the quasi-momenta using the spectral representation
\begin{equation}
  \begin{aligned}
    \hat p_l(x) &
    = \frac{\frac{x}{\kappa}\left(2\pi \chi\hat m_l+\hat \kappa_l\right)+2\pi \hat m_l}{\left(x-s\right)\left(x+s^{-1}\right)}
    +\int_{C_l} dy \frac{\hat \rho_l(y)}{x-y}
    +\int_{1/C_l} dy \frac{\hat {\tilde \rho}_l(y)}{x-y} ,
    \\
    \check p_l(x) &
    = \frac{\frac{x}{\kappa}\left(2 \pi \chi \check m_l-\check \kappa_l\right)-2\pi \check m_l}{\left(x+s\right)\left(x-s^{-1}\right)}
    +\int_{C_l} dy \frac{\check \rho_l(y)}{x-y}
    +\int_{1/C_l} dy \frac{\check {\tilde \rho}_l(y)}{x-y},
  \end{aligned}
\end{equation}
where $\rho_l(y)$ are the discontinuities at the cuts.
Thus, we have
\begin{equation}
  \begin{split}
    \hat p_l(1/x) &
    = \frac{\frac{x}{\kappa}\left(2 \pi \chi \hat m_l-\hat \kappa_l\right)-2\pi \hat m_l}{\left(x-s^{-1}\right)\left(x+s\right)}-2\pi \hat m_l \\
    &+\int_{1/C_l} dy \frac{\hat\rho_l(1/y)}{y}
    +\int_{C_l} dy \frac{\hat{\tilde \rho}_l(1/y)}{y}
    +\int_{1/C_l} dy \frac{\hat \rho_l(1/y)}{x-y}
    +\int_{C_l} dy \frac{\hat{\tilde \rho}_l(1/y)}{x-y},
  \end{split}
\end{equation}
and similarly for $\check p_l(1/x)$.

Up to this point the treatment is general for any direct sum of two identical supergroups. From now on we are going to assume a specific grading for the superalgebra as given in~\cite{Borsato:2013qpa}.
Writing $S=\sigma_1 \otimes \mathbb{S}_{lk}$ with $l,k=1,..,\mathrm{dim}(G_L)$,
and using~\eqref{eq:quasimomentasymm} we find that
\begin{equation}\label{eq:kappamconstraints}
  \begin{gathered}
    \hat \kappa_l = \mathbb{S}_{lm}\check \kappa_m,\quad
    \hat m_l = \mathbb{S}_{lm}\check m_m,
    \\
    2\pi \hat m_l =
    -\int_{C_l} dy \frac{\hat \rho_l(y)}{y}
    +\mathbb{S}_{lm}\int_{C_m} dy \frac{\check \rho_m(y)}{y}
    , \\
    2\pi \check m_l =
    +\int_{C_l} dy \frac{\check \rho_l(y)}{y}
    -\mathbb{S}_{lm}\int_{C_m} dy \frac{\hat \rho_m(y)}{y} ,
  \end{gathered}
\end{equation}
and
\begin{equation}
  \begin{aligned}
    \hat p_l(x) &
    = \frac{\frac{x}{\kappa}\left(2 \pi \chi \hat m_l+\hat \kappa_l\right)+2\pi \hat m_l}{\left(x-s\right)\left(x+s^{-1}\right)}
    +\int_{C_l} dy \frac{\hat \rho_l(y)}{x-y}
    -\mathbb{S}_{lm}\int_{C_m} \frac{dy}{y^2} \frac{\check \rho_m(y)}{x-y} \\
    \check p_l(x) &
    = \mathbb{S}_{lm}\frac{\frac{x}{\kappa}\left(2 \pi \chi \hat m_m-\hat \kappa_m\right)-2\pi \hat m_m}{\left(x+s\right)\left(x-s^{-1}\right)}
    +\int_{C_l} dy \frac{\check \rho_l(y)}{x-y}
    -\mathbb{S}_{lm}\int_{C_m} \frac{dy}{y^2} \frac{\hat \rho_m(y)}{x-y}.
  \end{aligned}
\end{equation}
Using~\eqref{eq:disc} we get
\begin{gather}
  \begin{aligned}
    2\pi \hat n_{k,i} =&
    \mathbb{A}_{kl}\frac{\frac{x}{\kappa}\left(2 \pi \chi \hat m_l+\hat \kappa_l\right)+2\pi \hat m_l}{\left(x-s\right)\left(x+s^{-1}\right)} \\
    &+\mathbb{A}_{kl}\pint_{C_l} dy \frac{\hat \rho_l(y)}{x-y}
    -\mathbb{A}_{kl}\mathbb{S}_{lm}\int_{C_m} \frac{dy}{y^2} \frac{\check \rho_m(y)}{x-1/y},
  \end{aligned} \\
  \begin{aligned}
    2\pi \check n_{k,i} =&
    -\mathbb{A}_{kl}\mathbb{S}_{lm}\frac{\frac{x}{\kappa}\left(2 \pi \chi \hat m_m-\hat \kappa_m\right)-2\pi \hat m_m}{\left(x-s^{-1}\right)\left(x+s\right)} \\
    &-\mathbb{A}_{kl}\pint_{C_l} dy \frac{\check \rho_l(y)}{x-y}
    +\mathbb{A}_{kl}\mathbb{S}_{lm}\int_{C_m} \frac{dy}{y^2} \frac{\hat \rho_m(y)}{x-1/y},
  \end{aligned}
\end{gather}
where we used
$A = \sigma_3 \otimes \mathbb{A}$.
We are going to use the $\algPSU(1,1|2)$ Cartan matrix
\begin{equation}
  \mathbb{A} =
  \begin{pmatrix}
    0 & -1 & 0 \\
    -1 & 2 & -1 \\
    0 & -1 & 0
  \end{pmatrix}
\end{equation}
and
\begin{equation}
  \mathbb{S} = 
  \begin{pmatrix}
    1  & -1 & 0 \\
    0  & -1 & 0 \\
    0  & -1 & 1
  \end{pmatrix} .
\end{equation}
The vectors $\hat{\kappa}$, $\hat{m}$ and $\check{m}$ are given by
\begin{equation}
  \begin{aligned}
    \hat \kappa & = 2\pi \mathcal{E} (1,0,1), \\
    \hat m & = -2(\hat {\mathcal{P}} - S \check {\mathcal{P}})
    = 2\left(
      -\hat {\mathcal{P}}_1+\check {\mathcal{P}}_1-\check {\mathcal{P}}_2,
      -\hat {\mathcal{P}}_2-\check {\mathcal{P}}_2,
      -\hat {\mathcal{P}}_3+\check {\mathcal{P}}_3-\check {\mathcal{P}}_2\right)
    \\
    \check m & = +2(\check {\mathcal{P}}- S \hat {\mathcal{P}})
    = 2\left( -\hat {\mathcal{P}}_1+\check {\mathcal{P}}_1+\hat {\mathcal{P}}_2,
      +\hat {\mathcal{P}}_2+\check {\mathcal{P}}_2,
      -\hat {\mathcal{P}}_3+\check {\mathcal{P}}_3+\hat {\mathcal{P}}_2\right) ,
  \end{aligned}
\end{equation}
with $\mathcal{P}_i = \frac{1}{4\pi} \int dy \frac{\rho_i(y)}{y}$. 
The level matching condition is $\hat {\mathcal{P}}_2+\check {\mathcal{P}}_2 = 0$.
Using these explicit matrices and vectors, the finite-gap equations are given by
\begin{align}
  2\pi \hat{n}_{1,i} &= - \int dy \frac{\hat{\rho}_2(y)}{x-y} - \int \frac{dy}{y^2} \frac{\check{\rho}_2(y)}{x - \frac{1}{y^2}} , \\
  \begin{split}
    2\pi \hat{n}_{2,i} &= 2 \pint dy \frac{\hat{\rho}_2(y)}{x-y} - \int dy \frac{\hat{\rho}_1(y)}{x-y} - \int dy \frac{\hat{\rho}_3(y)}{x-y}
    + \int \frac{dy}{y^2} \frac{\check{\rho}_1(y)}{x-\frac{1}{y}} + \int \frac{dy}{y^2} \frac{\check{\rho}_3(y)}{x-\frac{1}{y}} \\
    & \qquad - \frac{4\pi x}{(x-s)(x+s^{-1})} \bigl( \frac{1}{\kappa} \mathcal{E} - \frac{\chi}{\kappa} {\mathcal{M}} \bigr) + \frac{4\pi}{(x-s)(x+s^{-1})} {\mathcal{M}} ,
  \end{split} \\
  2\pi \hat{n}_{3,i} &= - \int dy \frac{\hat{\rho}_2(y)}{x-y} - \int \frac{dy}{y^2} \frac{\check{\rho}_2(y)}{x - \frac{1}{y^2}} , \\
  2\pi \check{n}_{1,i} &= + \int dy \frac{\check{\rho}_2(y)}{x-y} + \int \frac{dy}{y^2} \frac{\hat{\rho}_2(y)}{x - \frac{1}{y^2}} , \\
  \begin{split}
    2\pi \check{n}_{2,i} &= -2 \pint dy \frac{\check{\rho}_2(y)}{x-y} + \int dy \frac{\check{\rho}_1(y)}{x-y} + \int dy \frac{\check{\rho}_3(y)}{x-y}
    - \int \frac{dy}{y^2} \frac{\hat{\rho}_1(y)}{x-\frac{1}{y}} - \int \frac{dy}{y^2} \frac{\hat{\rho}_3(y)}{x-\frac{1}{y}} \\
    & \qquad - \frac{4\pi x}{(x+s)(x-s^{-1})} \bigl( \frac{1}{\kappa} \mathcal{E} + \frac{\chi}{\kappa} {\mathcal{M}} \bigr) + \frac{4\pi}{(x+s)(x-s^{-1})} {\mathcal{M}} ,
  \end{split} \\
  2\pi \check{n}_{3,i} &= + \int dy \frac{\check{\rho}_2(y)}{x-y} + \int \frac{dy}{y^2} \frac{\hat{\rho}_2(y)}{x - \frac{1}{y^2}} ,
\end{align}
where
\begin{equation}
  \mathcal{M} =  \hat {\mathcal{P}}_1
  -\check {\mathcal{P}}_1
  +2\check {\mathcal{P}}_2
  +\hat {\mathcal{P}}_3
  -\check {\mathcal{P}}_3.
\end{equation}
The integration contour is defined along the cuts associated with the density in the integrand, and $x\in \hat C_{l,i} / \check C_{l,i}$ in the $\hat n_{l,i} / \check n_{l,i}$ equations.


\section{Finite-gap equations from the Bethe equations}
\label{sec:finite-gap-from-BA}

In this section we will construct a set of Bethe equations based on the S-matrix presented in~\cite{Hoare:2013ida}, and compare the finite-gap limit of those equations with the construction in the previous section.

\subsection{The Hoare-Tseytlin S-matrix}
\label{sec:HT-S-matrix}

An S-matrix for the massive sector of light cone gauge string theory on $\AdS_3 \times \Sphere^3 \times \Torus^4$ with mixed RR and NSNS flux was proposed in~\cite{Hoare:2013ida}.
This S-matrix can be obtained from the S-matrix in the pure RR case~\cite{Borsato:2013qpa} by generalising the dispersion relation. After gauge fixing the superisometry algebra of the string theory background is broken to $\algPSU(1|1)^4 \ltimes \algU(1)^4$. Here the $\algU(1)^4$ factor indicates four central charges, which are related to the energy, momentum and the mass of the world-sheet excitations. The excitations form two irreducible representations of this algebra, and their dispersion relation are given by\footnote{%
  The coupling constant $h$ is related to the string tension $\lambda$ by $h = \frac{\sqrt{\lambda}}{4\pi} + \order(1)$. In the pure RR case it is known that there is no $\order(1)$ correction to $h$~\cite{Sundin:2012gc,Abbott:2012dd,Beccaria:2012kb}. We expect this to be true in the case of mixed fluxes as well~\cite{Sundin:2014}.
} %
\begin{equation}\label{eq:dispersion-relation}
  E = \sqrt{M^2 + 16h^2 \kappa^2 \sin^2\frac{p}{2}} .
\end{equation}
where $\kappa^2 = 1-\chi^2$ and the NSNS flux is proportional to $\chi$. In the $\chi = 0$ case the two representations both describe excitations of mass $M^2 = 1$. For non-zero $\chi$ the two masses are different. In~\cite{Hoare:2013lja} it was proposed that they take the form\footnote{%
A different set of masses was proposed in~\cite{Hoare:2013ida},
\begin{equation}
  M^2 = \bigl( 1 \pm 4h\chi\sin\frac{p}{2} \bigr)^2 .
\end{equation}
In the near-BMN limit, where $p \ll 1 \ll h$ with $hp$ kept fixed, the two mass terms both give
\begin{equation}
  M^2 \approx ( 1 \pm 2\chi h p )^2 + \dotsb ,
\end{equation}
in agreement with the tree-level calculation of~\cite{Hoare:2013pma}. The finite-gap limit is only sensitive to the near-BMN part of the dispersion relation so we are free to use any of the two expressions above for the mass.
}%
\begin{equation}
  \hat{M}^2 = ( 1 + 2\chi h p )^2 , \qquad
  \check{M}^2 = ( 1 - 2\chi h p )^2 .
\end{equation}
The all-loop S-matrix of~\cite{Borsato:2013qpa} is expressed in terms of a set of spectral parameters $x^\pm$ which are parametrizing the central charges carried by the excitations. For $\chi \neq 0$ there are excitations with two different masses. We will denote the corresponding spectral parameters by $\hat{x}^{\pm}$ and $\check{x}^{\pm}$. They are related to the world-sheet momentum by
\begin{equation}
  \frac{\hat{x}^+}{\hat{x}^-} = e^{+ip} , \qquad
  \frac{\check{x}^+}{\check{x}^-} = e^{+ip} .
\end{equation}
The spectral parameters further satisfy shortening conditions of the form
\begin{equation}\label{eq:shortening-condition}
  \begin{aligned}
    \hat{x}^+ + \frac{1}{\hat{x}^+} - \hat{x}^- - \frac{1}{\hat{x}^-} 
    &= \frac{i \hat{M}}{\kappa h} = \frac{i}{\kappa h} \bigl( 1 + 2 \frac{\chi}{\kappa} \hat{P} \bigr) , \\
    \check{x}^+ + \frac{1}{\check{x}^+} - \check{x}^- - \frac{1}{\check{x}^-} 
    &= \frac{i \check{M}}{\kappa h} = \frac{i}{\kappa h} \bigl( 1 - 2 \frac{\chi}{\kappa} \check{P} \bigr) ,
  \end{aligned}
\end{equation}
where
\begin{equation}
  \hat{P} = \kappa h p , \qquad
  \check{P} = \kappa h p .
\end{equation}
The dispersion relations~\eqref{eq:dispersion-relation} can be expressed in terms of $\hat{x}^\pm$ and $\check{x}^\pm$ as
\begin{equation}\label{eq:dispersion-relation-x}
  \begin{aligned}
    \hat{E} &= -i \kappa h \Bigl( \Bigl( \hat{x}^+ - \hat{x}^- \Bigl) - \Bigl( \frac{1}{\hat{x}^+} - \frac{1}{\hat{x}^-} \Bigr) \Bigr) ,
    \\
    \check{E} &= -i \kappa h \Bigl( \Bigl( \check{x}^+ - \check{x}^- \Bigl) - \Bigl( \frac{1}{\check{x}^+} - \frac{1}{\check{x}^-} \Bigr) \Bigr) .
  \end{aligned}
\end{equation}
To expand this in the near BMN limit~\cite{Berenstein:2002jq} we take $h$ large and keep $\hat{P}$ and $\check{P}$ fixed.

The matrix elements of the S-matrix of~\cite{Hoare:2013ida} takes the same form as those in~\cite{Borsato:2013qpa}, but expressed in terms of the deformed spectral parameters. The S-matrix also contains scalar factors that give the overall normalisation. Since the massive excitations transform in two different representations of the symmetry algebra, there are four such scalar factors -- one for each combination of two representation. However, the phases are pairwise related by discrete symmetries and can be expressed in terms of two different functions. The phase $\sigma(p_1,p_2)$ gives the phase when scattering two excitations from the same multiplet, while the phase for scattering excitations from different representations is given by $\bar{\sigma}(p_1,p_2)$. The two phases are related by the crossing equations~\cite{Borsato:2013qpa,Borsato:2013hoa}\footnote{%
  Due to the momentum dependence on the right hand side of the shortening conditions~\eqref{eq:shortening-condition}, the analytical structure of the spectral parameters is more complicated for $\chi \neq 0$ than for $\chi = 0$ and it is not obvious how to the analytical continuation in the crossing equations should be performed. However, in this paper we only need the leading part of the phases at strong coupling, and in that limit the crossing equations take the same form for any $\chi$.%
}%
\begin{align}
  \label{eq:crossing-equation-1}
  \sigma^2(x^\pm,y^\pm)\,\bar{\sigma}^2(x^\pm,1/y^\pm) &= \left(\frac{x^+}{x^-}\right)^2\frac{(x^--y^+)^2}{(x^--y^-)(x^+-y^+)}\frac{1-\frac{1}{x^-y^+}}{1-\frac{1}{x^+y^-}},
  \\
  \label{eq:crossing-equation-2}
  \sigma^2(x^\pm,1/\bar{y}^\pm)\,\bar{\sigma}^2(x^\pm,\bar{y}^\pm) &= \left(\frac{x^+}{x^-}\right)^2\frac{\left(1-\frac{1}{x^-\bar{y}^-}\right)\left(1-\frac{1}{x^+\bar{y}^+}\right)}{\left(1-\frac{1}{x^+\bar{y}^-}\right)^2}\frac{x^--\bar{y}^+}{x^+-\bar{y}^-} .
\end{align}
Here $x^\pm$ and $y^\pm$ indicate two spectral parameters of the same kind (either $\hat{x}^\pm$ or $\check{x}^\pm$), while $\bar{y}^\pm$ is a spectral parameter of the opposite kind ($\check{x}^\pm$ or $\hat{x}^\pm$).
As we will see later some care is needed in order to find a good form for these scalar factors in the finte gap limit.

\subsection{Bethe ansatz equations}
\label{sec:Bethe-ansatz}

The Bethe equations in the pure RR case were constructed in~\cite{Borsato:2013qpa} by diagonalising the S-matrix. Since the S-matrix takes the same form when expressed in term of the spectral parameters also for mixed fluxes, the same construction is valid in that case as well. For completeness we write out these equations here. In total there are six equations. There are two sets of momentum carrying Bethe roots $\hat{x}_{2,k}^\pm$ and $\check{x}_{2,k}^\pm$, with $k=1,\dotsc,\hat{K}_2$ and $k=1,\dotsc,\check{K}_2$. In addition there are four sets of auxiliary roots $\hat{x}_{1,k}$, $\hat{x}_{3,k}$, $\check{x}_{1,k}$ and $\check{x}_{3,k}$. We denote the corresponding excitation  numbers by $\hat{K}_1$, $\hat{K}_3$, $\check{K}_1$ and $\check{K}_3$.

The Bethe equations take the form\footnote{%
  The pure RR flux Bethe equations can be constructed from an integrable spin-chain~\cite{OhlssonSax:2011ms,Borsato:2013qpa}. 
  The linear dependence of $p$ in $\hat{M}$ and $\check{M}$ discussed in the last section makes the dispersion relation in the mixed flux case non-periodic in the momentum.
  Hence, a spin-chain interpretation of the Bethe ansatz equations does not seem very natural here.
}%
\begin{align}
    \label{eq:BA-1a}
    1 &=
    \prod_{j=1}^{\hat{K}_2} \frac{\hat{x}_{1,k} - \hat{x}_{2,j}^+}{\hat{x}_{1,k} - \hat{x}_{2,j}^-}
    \prod_{j=1}^{\check{K}_2} \frac{1 - \frac{1}{\hat{x}_{1,k} \check{x}_{2,j}^-}}{1- \frac{1}{\hat{x}_{1,k} \check{x}_{2,j}^+}} , \\
    \begin{split}
    \label{eq:BA-2a}
      \left(\frac{\hat{x}_{2,k}^+}{\hat{x}_{2,k}^-}\right)^L &=
      \prod_{\substack{j = 1\\j \neq k}}^{\hat{K}_2} \frac{\hat{x}_{2,k}^+ - \hat{x}_{2,j}^-}{\hat{x}_{2,k}^- - \hat{x}_{2,j}^+} \frac{1- \frac{1}{\hat{x}_{2,k}^+ \hat{x}_{2,j}^-}}{1- \frac{1}{\hat{x}_{2,k}^- \hat{x}_{2,j}^+}} \sigma^2(\hat{x}_{2,k},\hat{x}_{2,j})
      \prod_{j=1}^{\hat{K}_1} \frac{\hat{x}_{2,k}^- - \hat{x}_{1,j}}{\hat{x}_{2,k}^+ - \hat{x}_{1,j}}
      \prod_{j=1}^{\hat{K}_3} \frac{\hat{x}_{2,k}^- - \hat{x}_{3,j}}{\hat{x}_{2,k}^+ - \hat{x}_{3,j}}
      \\ &\,\times
      \prod_{j=1}^{\check{K}_2} \frac{1- \frac{1}{\hat{x}_{2,k}^+ \check{x}_{2,j}^+}}{1- \frac{1}{\hat{x}_{2,k}^- \check{x}_{2,j}^-}} \frac{1- \frac{1}{\hat{x}_{2,k}^+ \check{x}_{2,j}^-}}{1- \frac{1}{\hat{x}_{2,k}^- \check{x}_{2,j}^+}} \bar{\sigma}^2(\hat{x}_{2,k},\check{x}_{2,j})
      \prod_{j=1}^{\check{K}_1} \frac{1 - \frac{1}{\hat{x}_{2,k}^- \check{x}_{1,j}}}{1- \frac{1}{\hat{x}_{2,k}^+ \check{x}_{1,j}}}
      \prod_{j=1}^{\check{K}_3} \frac{1 - \frac{1}{\hat{x}_{2,k}^- \check{x}_{3,j}}}{1- \frac{1}{\hat{x}_{2,k}^+ \check{x}_{3,j}}} ,
    \end{split} \\
    \label{eq:BA-3a}
    1 &=
    \prod_{j=1}^{\hat{K}_2} \frac{\hat{x}_{3,k} - \hat{x}_{2,j}^+}{\hat{x}_{3,k} - \hat{x}_{2,j}^-}
    \prod_{j=1}^{\check{K}_2} \frac{1 - \frac{1}{\hat{x}_{3,k} \check{x}_{2,j}^-}}{1- \frac{1}{\hat{x}_{3,k} \check{x}_{2,j}^+}} , \\
    \label{eq:BA-1b}
    1 &=
    \prod_{j=1}^{\check{K}_2} \frac{\check{x}_{1,k} - \check{x}_{2,j}^-}{\check{x}_{1,k} - \check{x}_{2,j}^+}
    \prod_{j=1}^{\hat{K}_2} \frac{1 - \frac{1}{\check{x}_{1,k} \hat{x}_{2,j}^+}}{1- \frac{1}{\check{x}_{1,k} \hat{x}_{2,j}^-}} , \\
    \begin{split}
    \label{eq:BA-2b}
      \left(\frac{\check{x}_{2,k}^+}{\check{x}_{2,k}^-}\right)^L &=
      \prod_{\substack{j = 1\\j \neq k}}^{\check{K}_2} \frac{\check{x}_{2,k}^- - \check{x}_{2,j}^+}{\check{x}_{2,k}^+ - \check{x}_{2,j}^-} \frac{1- \frac{1}{\check{x}_{2,k}^+ \check{x}_{2,j}^-}}{1- \frac{1}{\check{x}_{2,k}^- \check{x}_{2,j}^+}} \sigma^2(\check{x}_{2,k},\check{x}_{2,j})
      \prod_{j=1}^{\check{K}_1} \frac{\check{x}_{2,k}^+ - \check{x}_{1,j}}{\check{x}_{2,k}^- - \check{x}_{1,j}}
      \prod_{j=1}^{\check{K}_3} \frac{\check{x}_{2,k}^+ - \check{x}_{3,j}}{\check{x}_{2,k}^- - \check{x}_{3,j}}
      \\ &\,\times
      \prod_{j=1}^{\hat{K}_2} \frac{1- \frac{1}{\check{x}_{2,k}^- \hat{x}_{2,j}^-}}{1- \frac{1}{\check{x}_{2,k}^+ \hat{x}_{2,j}^+}} \frac{1- \frac{1}{\check{x}_{2,k}^+ \hat{x}_{2,j}^-}}{1- \frac{1}{\check{x}_{2,k}^- \hat{x}_{2,j}^+}} \bar{\sigma}^2(\check{x}_{2,k},\hat{x}_{2,j})
      \prod_{j=1}^{\hat{K}_{1}} \frac{1 - \frac{1}{\check{x}_{2,k}^+ \hat{x}_{1,j}}}{1- \frac{1}{\check{x}_{2,k}^- \hat{x}_{1,j}}}
      \prod_{j=1}^{\hat{K}_{3}} \frac{1 - \frac{1}{\check{x}_{2,k}^+ \hat{x}_{3,j}}}{1- \frac{1}{\check{x}_{2,k}^- \hat{x}_{3,j}}} ,
    \end{split} \\
    \label{eq:BA-3b}
    1 &=
    \prod_{j=1}^{\check{K}_2} \frac{\check{x}_{3,k} - \check{x}_{2,j}^-}{\check{x}_{3,k} - \check{x}_{2,j}^+}
    \prod_{j=1}^{\hat{K}_2} \frac{1 - \frac{1}{\check{x}_{3,k} \hat{x}_{2,j}^+}}{1- \frac{1}{\check{x}_{3,k} \hat{x}_{2,j}^-}} .
\end{align}
The Bethe roots further satisfy the level matching condition
\begin{equation}
  \prod_{i=k}^{\hat{K}_2} \frac{\hat{x}^+_{2,k}}{\hat{x}^-_{2,k}} \prod_{i=k}^{\check{K}_2} \frac{\check{x}^+_{2,k}}{\check{x}^-_{2,k}} = 1 .
\end{equation}
The local charges $\mathcal{Q}_n$ are given by
\begin{equation}
  \mathcal{Q}_n = \sum_{k=1}^{\hat{K}_2} \hat{\mathcal{Q}}_n(\hat{x}_{2,k}^\pm) + \sum_{k=1}^{\check{K}_2} \check{\mathcal{Q}}_n(\check{x}_{2,k}^\pm) ,
\end{equation}
where each Bethe root gives a contribution
\begin{equation}
  \begin{aligned}
    \hat{\mathcal{Q}}_n(\hat{x}^\pm) &= \frac{i}{n-1} \biggl(\frac{1}{(\hat{x}^+)^{n-1}} - \frac{1}{(\hat{x}^-)^{n-1}} \biggr) , \\
    \check{\mathcal{Q}}_n(\check{x}^\pm) &= \frac{i}{n-1} \biggl(\frac{1}{(\check{x}^+)^{n-1}} - \frac{1}{(\check{x}^-)^{n-1}} \biggr) .
  \end{aligned}
\end{equation}

We will assume that the dressing phase $\sigma$ to the leading order is given by the Arutyunov--Frolov--Staudacher (AFS) phase~\cite{Arutyunov:2004vx}\footnote{%
  In equation~(5.29) of~\cite{Hoare:2013ida} a different form of the AFS phase was given for the mixed flux S-matrix.
  For $\chi=0$ that expression is equivalent to~\eqref{eq:AFS-phase} provided we impose the shortening condition satisfied by $x^\pm$. 
  For non-zero $\chi$ the shortening condition is deformed in a momentum dependent way, as shown in equation~\eqref{eq:shortening-condition}, and the two forms for the AFS phase is no longer equal.
    However, the expressions differ only at higher orders of the large $h$ expansion, so either form can be used to reproduced the tree-level S-matrix of~\cite{Hoare:2013pma}.
}%
\begin{equation}
  - \frac{i}{\kappa h} \log \sigma_{\text{AFS}}(x^\pm,y^\pm) = \chi(x^+,y^+) - \chi(x^+,y^-) - \chi(x^-,y^+) + \chi(x^-,y^-) ,
\end{equation}
where
\begin{equation}\label{eq:AFS-phase}
  \chi(x,y) = \left( y + \frac{1}{y} - x - \frac{1}{x} \right) \log \left( 1 - \frac{1}{xy} \right) .
\end{equation}
The phase $\bar{\sigma}$ can then be obtained from the crossing relation~\eqref{eq:crossing-equation-2}.\footnote{%
  The crossing equations describe how the phases $\sigma$ and $\bar{\sigma}$ behave under analytical continuation outside the physical region. Hence they relate the two phases on different sheets of a Riemann surface, and can not be used to express the phase $\bar{\sigma}$ in terms of $\sigma$. However, to the leading order at strong coupling the functions can trivially be continued back to the physical region, so in this limit it is enough to give an expression for one of the phases.%
} %
The resulting phase can be written in an expansion in local charges as
\begin{equation}\label{eq:sigma-tilde-sum}
  \begin{split}
    - \frac{i}{\kappa h} \log \bar{\sigma} (\hat{x}^\pm,\check{y}^\pm) = 
    + &\sum_{n=2}^{\infty} \Bigl(
    \hat{\mathcal{Q}}_n (\hat{x}^\pm) \check{\mathcal{Q}}_{n+1} (\check{y}^\pm) - \hat{\mathcal{Q}}_{n+1} (\hat{x}^\pm)  \check{\mathcal{Q}}_n (\check{y}^\pm)
    \Bigr) \\
    + & \frac{\chi}{\kappa} \Bigl(
    \hat{\mathcal{Q}}_1 (\hat{x}^\pm) \check{\mathcal{Q}}_1 (\check{y}^\pm)
    - 2 \sum_{n=2}^{\infty} \hat{\mathcal{Q}}_n (\hat{x}^\pm) \check{\mathcal{Q}}_n (\check{y}^\pm)
    \Bigr) .
  \end{split}
\end{equation}
The first line here is the same as for the AFS phase, but the second line contains two new features. Firstly, it contains the first charges $\hat{\mathcal{Q}}_1$ and $\check{\mathcal{Q}}_1$, while the AFS phase only involves charges $\hat{\mathcal{Q}}_n$ and $\check{\mathcal{Q}}_n$ with $n \ge 2$. Secondly, the second line is symmetric under the exchange of $\hat{x}^{\pm}$ and $\check{y}^{\pm}$. This is not in contradiction with unitarity. The inverse scattering phase is obtained by additionally sending $\chi \to -\chi$ which gives an additional minus sign in the second line.
By performing the sums we again obtain an expansion of the form
\begin{equation}
  - \frac{i}{\kappa h} \log \bar{\sigma}(\hat{x}^\pm,\check{y}^\pm) = 
  \bar{\chi}(\hat{x}^+,\check{y}^+) - \bar{\chi}(\hat{x}^+,\check{y}^-) 
  - \bar{\chi}(\hat{x}^-,\check{y}^+) + \bar{\chi}(\hat{x}^-,\check{y}^-) .
\end{equation}
The function $\bar{\chi}(\hat{x},\check{y})$ takes the form
\begin{equation}
  \bar{\chi}(\hat{x},\check{y}) = 
  \Bigl( \check{y} + \frac{1}{\check{y}} - \hat{x} - \frac{1}{\hat{x}} \Bigr) \log \Bigl( 1 - \frac{1}{\hat{x}\check{y}} \Bigr) 
  - \frac{\chi}{\kappa} \Bigl( 2 \operatorname{Li}_2 \Bigl( \frac{1}{\hat{x}\check{y}} \Bigr) - \log \hat{x} \, \log \check{y} \Bigr) ,
\end{equation}
where $\operatorname{Li}_2$ is the dilogarithm.

By expanding the Bethe equations at large $x$ we can obtain the global charges of the corresponding solution. Denoting the angular momenta on $\Sphere^3$ by $J$ and $K$, the angular momentum on $\AdS_3$ by $S$ and the global AdS energy by $D$ we find
\begin{align}
  D &= + \check{K}_2 + \frac{1}{2} \bigl( \hat{K}_1 + \hat{K}_3 - \check{K}_1 - \check{K}_3 \bigr) + L + \delta D , \\
  J &= - \hat{K}_2 + \frac{1}{2} \bigl( \hat{K}_1 + \hat{K}_3 - \check{K}_1 - \check{K}_3 \bigr) + L , \\
  K &= - \hat{K}_2 + \frac{1}{2} \bigl( \hat{K}_1 + \hat{K}_3 + \check{K}_1 + \check{K}_3 \bigr) - 2 \frac{\chi}{\kappa} ( \hat{P} + \check{P} ) , \\
  S &= - \check{K}_2 + \frac{1}{2} \bigl( \hat{K}_1 + \hat{K}_3 + \check{K}_1 + \check{K}_3 \bigr) .
\end{align}
In deriving these expressions there is an essential contribution from the phase $\bar{\sigma}$, which arises due to the presence of the charges $\hat{\mathcal{Q}}_1$ and $\check{\mathcal{Q}}_1$ in~\eqref{eq:sigma-tilde-sum}. Without this term the global charges would depend separately on $\hat{P}$ and $\check{P}$. In the above expression for the charges these parameters only appear in a combination proportional to the total world-sheet momentum $p$. For a physical state we have $p = 2\pi m$ for some integer $m$. Moreover, quantization of the WZ coupling can be expressed as $4\pi h \chi = k$ with $k \in \Integers$~\cite{Cagnazzo:2012se,Hoare:2013lja}. Hence the combination
\begin{equation}
  2 \frac{\chi}{\kappa} ( \hat{P} + \check{P} ) = 2 \chi h p = m k
\end{equation}
is also an integer, and the angular momentum $K$ is quantized as expected.

The \emph{anomalous dimension} is given by
\begin{equation}
  \delta D =
  2\kappa h \mathcal{Q}_2 + 2 \frac{\chi}{\kappa} ( \hat{P} - \check{P} ) .
\end{equation}
This gives the worldsheet Hamiltonian
\begin{equation}\label{eq:energy-KKdD}
  D - J = \hat{K}_2 + \check{K}_2 + \delta D .
\end{equation}
Since the dispersion relations of the excitations are given by~\eqref{eq:dispersion-relation-x}, the total energy can also be written as
\begin{equation}\label{eq:energy-Ek}
  D - J = \sum_{k=1}^{\hat{K}_2} \hat{E}_k + \sum_{k=1}^{\check{K}_2} \check{E}_k .
\end{equation}
To see that the above expressions are equal we can use the shortening conditions~\eqref{eq:shortening-condition} to rewrite the sums above. For the first sum this gives
\begin{align}
  \sum_{k=1}^{\hat{K}_2} \hat{E}_k
  &= - i\kappa h \sum_{k=1}^{\hat{K}_2} \biggl( \biggl( \hat{x}_{2,k}^+ - \hat{x}_{2,k}^- \biggr) - \biggl( \frac{1}{\hat{x}_{2,k}^+} - \frac{1}{\hat{x}_{2,k}^+} \biggr) \biggr) \\
  &= + \sum_{k=1}^{\hat{K}_2} \biggl( 2 i\kappa h \biggl( \frac{1}{\hat{x}_{2,k}^+} - \frac{1}{\hat{x}_{2,k}^+} \biggr) + 1 + 2 \frac{\chi}{\kappa} \hat{P} \biggr)
  = \hat{K}_2 + \delta\hat{D} ,
\end{align}
where $\delta \hat{D}$ is the contribution to the charge $\delta D$ from the $\hat{K}_2$ roots. Rewriting the second sum in a similar way we see that~\eqref{eq:energy-Ek} and~\eqref{eq:energy-KKdD} agree.

Another useful combination of charges is
\begin{equation}
  M = S - K = \hat{K}_2 - \check{K}_2 + 2\frac{\chi}{\kappa} ( \hat{P} + \check{P} ) ,
\end{equation}
which gives the mass $M^2$ appearing in the dispersion relations~\eqref{eq:dispersion-relation-x}.

\subsection{Scaling limit of the Bethe equations}
\label{sec:finite-gap-limit}

To obtain the finite-gap equations we consider a solution to the Bethe equations with large angular momentum and a large number of excitations, so that $L \approx K_i \gg 1$. We further take the strong coupling limit $h \gg 1$. The Bethe roots then condense and form a set of cuts in the complex plane~\cite{Kazakov:2004qf}. We will denote the collection of cuts which arise from the condensation of the roots $\hat{x}_{i,k}$ and $\check{x}_{i,k}$ by $\hat{C}_i$ and $\check{C}_i$, respectively.

At large coupling we can solve the shortening conditions~\eqref{eq:shortening-condition} by setting
\begin{equation}
  \label{eq:x-expansion}
  \hat{x}^\pm = x \pm \frac{i}{2} \hat{\alpha}(x) + \order(1/h^2), \qquad
  \check{x}^\pm = x \pm \frac{i}{2} \check{\alpha}(x) + \order(1/h^2) ,
\end{equation}
where we have introduced the functions
\begin{equation}\label{eq:alphas}
  \hat{\alpha}(x) = \frac{1}{\kappa h} \frac{x^2}{(x-s)(x+s^{-1})} , \qquad
  \check{\alpha}(x) = \frac{1}{\kappa h} \frac{x^2}{(x+s)(x-s^{-1})} ,
\end{equation}
and
\begin{equation}
  s = \sqrt{\frac{1+\chi}{1-\chi}}.
\end{equation}
We further introduce the densities
\begin{equation}
  \hat{\rho}_i(x) = \sum_k \alpha(\hat{x}_{i,k}) \delta(x-\hat{x}_{i,k}) , \qquad
  \check{\rho}_i(x) = \sum_k \check{\alpha}(\check{x}_{i,k}) \delta(x-\check{x}_{i,k})
\end{equation}
along the cuts formed by the condensed Bethe roots. The finite-gap equations can now be found by taking the logarithm of the Bethe equations in the above limit. This gives
\begin{align}
  2\pi \hat{n}_1 &= - \int dy \frac{\hat{\rho}_2(y)}{x-y} - \int \frac{dy}{y^2} \frac{\check{\rho}_2(y)}{x - \frac{1}{y^2}} , \label{eq:finite-gap-1a}\\
  \begin{split}
    2\pi \hat{n}_2 &= 2 \pint dy \frac{\hat{\rho}_2(y)}{x-y} - \int dy \frac{\hat{\rho}_1(y)}{x-y} - \int dy \frac{\hat{\rho}_3(y)}{x-y}
    + \int \frac{dy}{y^2} \frac{\check{\rho}_1(y)}{x-\frac{1}{y}} + \int \frac{dy}{y^2} \frac{\check{\rho}_3(y)}{x-\frac{1}{y}} \\
    & \qquad - \frac{4\pi x}{(x-s)(x+s^{-1})} \bigl( \frac{1}{\kappa} \mathcal{E} - \frac{\chi}{\kappa} \mathcal{M} \bigr) + \frac{4\pi}{(x-s)(x+s^{-1})} \mathcal{M} ,
  \end{split} \label{eq:finite-gap-2a} \\
  2\pi \hat{n}_3 &= - \int dy \frac{\hat{\rho}_2(y)}{x-y} - \int \frac{dy}{y^2} \frac{\check{\rho}_2(y)}{x - \frac{1}{y^2}} , \label{eq:finite-gap-3a} \\
  2\pi \check{n}_1 &= + \int dy \frac{\check{\rho}_2(y)}{x-y} + \int \frac{dy}{y^2} \frac{\hat{\rho}_2(y)}{x - \frac{1}{y^2}} , \label{eq:finite-gap-1b} \\
  \begin{split}
    2\pi \check{n}_2 &= -2 \pint dy \frac{\check{\rho}_2(y)}{x-y} + \int dy \frac{\check{\rho}_1(y)}{x-y} + \int dy \frac{\check{\rho}_3(y)}{x-y}
    - \int \frac{dy}{y^2} \frac{\hat{\rho}_1(y)}{x-\frac{1}{y}} - \int \frac{dy}{y^2} \frac{\hat{\rho}_3(y)}{x-\frac{1}{y}} \\
    & \qquad - \frac{4\pi x}{(x+s)(x-s^{-1})} \bigl( \frac{1}{\kappa} \mathcal{E} + \frac{\chi}{\kappa} \mathcal{M} \bigr) + \frac{4\pi}{(x+s)(x-s^{-1})} \mathcal{M} ,
  \end{split} \label{eq:finite-gap-2b} \\
  2\pi \check{n}_3 &= + \int dy \frac{\check{\rho}_2(y)}{x-y} + \int \frac{dy}{y^2} \frac{\hat{\rho}_2(y)}{x - \frac{1}{y^2}} . \label{eq:finite-gap-3b}
\end{align}
The integrals above should be taken over the cuts along which the densities in the integrand are defined.
The parameters $x$ appearing on the right hand sides of the above equations take values along the cuts, so that the first (last) three equations should be evaluated for $x$ on $\hat{C}_1$, $\hat{C}_2$ and $\hat{C}_3$ ($\check{C}_1$, $\check{C}_2$ and $\check{C}_3$), respectively. Note that this means that the first and third (and fourth and sixth) are not actually identical even though they take the same form as written here.

In order to write down the coefficients $\mathcal{E}$ and $\mathcal{M}$ we introduce
\begin{equation}
  \begin{aligned}
    \hat{\mathcal{P}}_m &= \frac{1}{4\pi} \int \frac{dy}{y} \hat{\rho}_m(y) , \qquad &
    \hat{\mathcal{E}}_m &= \frac{\kappa}{4\pi} \int \frac{dy}{y^2} \hat{\rho}_m(y) , \\
    \check{\mathcal{P}}_m &= \frac{1}{4\pi} \int \frac{dy}{y} \check{\rho}_m(y) , \qquad &
    \check{\mathcal{E}}_m &= \frac{\kappa}{4\pi} \int \frac{dy}{y^2} \check{\rho}_m(y) .
  \end{aligned}
\end{equation}
The residues appearing in the finte gap equations are then given by
\begin{equation}
  \mathcal{M} = + \hat{\mathcal{P}}_1 + \hat{\mathcal{P}}_3 - \check{\mathcal{P}}_1 + 2\check{\mathcal{P}}_2 - \check{\mathcal{P}}_3 ,
\end{equation}
and
\begin{equation}
  \mathcal{E} = \mathcal{L} - \hat{\mathcal{E}}_1 + 2\hat{\mathcal{E}}_2 - \hat{\mathcal{E}}_3 + \check{\mathcal{E}}_1 + \check{\mathcal{E}}_3
  - \chi \bigl(
  \hat{\mathcal{P}}_1 - 2\hat{\mathcal{P}}_2 + \hat{\mathcal{P}}_3 + \check{\mathcal{P}}_1  + \check{\mathcal{P}}_3
  \bigr) ,
\end{equation}
with $\mathcal{L} = L / \sqrt{\lambda}$.
The finite-gap equations obtained from the Bethe ansatz agree with the results of section~\ref{sec:super-coset-B-field-integrability}.
For $s = 1$ the above expressions exactly agree with the results of~\cite{Borsato:2013qpa}, as expected. The anomalous dimension is given by
\begin{equation}
  \frac{\delta D}{\sqrt{\lambda}} = 2 (\hat{\mathcal{E}}_2 + \check{\mathcal{E}}_2) + 2 \chi ( \hat{\mathcal{P}}_2 - \check{\mathcal{P}}_2 ) ,
\end{equation}
while the total worldsheet momentum is\footnote{%
  The factor $4\pi$ appears because we are taking the finite-gap limit by scaling the momentum by $1/ \kappa h$ rather than by $1/\sqrt{\lambda}$.%
} %
\begin{equation}
  p_{\text{total}} = 4\pi(\hat{\mathcal{P}}_2 + \check{\mathcal{P}}_2) .
\end{equation}
For a physical string state $p_{\text{total}} \in 2\pi \Integers$.


\section{Quasi-momenta in the fundamental representation}
\label{sec:quasi-momenta}

In this section we write down the quasi-momenta in an explicit representation. Let us introduce the resolvents
\begin{equation}
G_{\tilde{a}}(x) = \sum_{k=1}^{K_{\tilde{a}}}\frac{\tilde \alpha(x_{\tilde{a},k})}{x-x_{\tilde{a},k}},\quad
H_{\tilde{a}}(x) = \sum_{k=1}^{K_{\tilde{a}}}\frac{\tilde \alpha(x)}{x-x_{\tilde{a},k}},\quad
\bar G_{\tilde{a}}(x) = G_{\tilde{a}} (1/x),\quad
\bar H_{\tilde{a}} (x) = H_{\tilde{a}} (1/x),
\end{equation}
where tilde stands for hat or check. These are related to the integrals appearing in the finite-gap equations by the relations
\begin{gather}
  \int dy \frac{\hat{\rho}_a(y)}{x-y}
  = + G_{\hat{a}}
  = + H_{\hat{a}} - \frac{G_{\hat{a}}(0) + x G_{\hat{a}}'(0) + 2x\frac{\chi}{\kappa} G_{\hat{a}}(0)}{(x-s)(x+s^{-1})} ,
  \\
  \int dy \frac{\check{\rho}_a(y)}{x-y}
  = + G_{\check{a}}
  = + H_{\check{a}} - \frac{G_{\check{a}}(0) + x G_{\check{a}}'(0) - 2x\frac{\chi}{\kappa} G_{\check{a}}(0)}{(x+s)(x-s^{-1})} ,
  \\
  \int \frac{dy}{y^2} \frac{\hat{\rho}_a(y)}{x-\frac{1}{y}}
  = - \bar{G}_{\hat{a}} + G_{\hat{a}}(0)
  = - \bar{H}_{\hat{a}} - \frac{G_{\hat{a}}(0) + x G_{\hat{a}}'(0)}{(x+s)(x-s^{-1})}
  \\
  \int \frac{dy}{y^2} \frac{\check{\rho}_a(y)}{x-\frac{1}{y}}
  = - \bar{G}_{\check{a}} + G_{\check{a}}(0)
  = - \bar{H}_{\check{a}} - \frac{G_{\check{a}}(0) + x G_{\check{a}}'(0)}{(x-s)(x+s^{-1})}
  \\
  \tilde{G}_a(0) = -4\pi \tilde{\mathcal{P}}_a , 
  \qquad
  \tilde{G}_a(0)' = -\frac{4\pi}{\kappa} \tilde{\mathcal{E}}_a , 
\end{gather}
Inserting this into the finite-gap equations we find
\begin{equation}
   \begin{aligned}\label{eq:fundQM}
    2\pi n_{\hat 1}  = & -H_{\hat 2} +\bar H_{\check 2} + \frac{G_{\hat 2}(0)\left(1+2\frac{\chi}{\kappa}x\right) +G_{\hat 2}'(0)x}{(x-s)(x+s^{-1})} +\frac{G_{\check 2}(0) +G_{\check 2}'(0)x}{(x-s)(x+s^{-1})}, \\
    2\pi n_{\hat 2} = & +2H_{\hat 2} -H_{\hat 1} -H_{\hat 3} -\bar H_{\check 1} -\bar H_{\check 3}  \\
    &-2\frac{G_{\hat 2}(0) +G_{\check 2}(0)}{(x-s)(x+s^{-1})}\left(1+\frac{\chi}{\kappa}x\right)-\frac{\frac{4\pi}{\kappa}\mathcal{L} x}{(x-s)(x+s^{-1})}, \\
    2\pi n_{\hat 3} = & -H_{\hat 2} +\bar H_{\check 2} + \frac{G_{\hat 2}(0)\left(1+2\frac{\chi}{\kappa}x\right) +G_{\hat 2}'(0)x}{(x-s)(x+s^{-1})} +\frac{G_{\check 2}(0) +G_{\check 2}'(0)x}{(x-s)(x+s^{-1})}, \\
    2\pi n_{\check 1} = & +H_{\check 2} -\bar H_{\hat 2} -\frac{G_{\check 2}(0)\left(1-2\frac{\chi}{\kappa}x\right) +G_{\check 2}'(0)x}{(x+s)(x-s^{-1})} -\frac{G_{\hat 2}(0) +G_{\hat 2}'(0)x}{(x+s)(x-s^{-1})}, \\
    2\pi n_{\check 2} = & -2H_{\check 2} +H_{\check 1} +H_{\check 3} +\bar H_{\hat 1} +\bar H_{\hat 3} \\
    &+2 x \frac{\frac{\chi}{\kappa}\left(G_{\hat 2}(0)-G_{\check 2}(0)\right) +\left(G_{\hat 2}'(0)+G_{\check 2}'(0)\right)}{(x+s)(x-s^{-1})}
    -\frac{\frac{4\pi}{\kappa}\mathcal{L} x}{(x+s)(x-s^{-1})}, \\
    2\pi n_{\check 3} = & +H_{\check 2} -\bar H_{\hat 2} -\frac{G_{\check 2}(0)\left(1-2\frac{\chi}{\kappa}x\right) +G_{\check 2}'(0)x}{(x+s)(x-s^{-1})} -\frac{G_{\hat 2}(0) +G_{\hat 2}'(0)x}{(x+s)(x-s^{-1})} .
  \end{aligned}
\end{equation}
If we define the eight quasi-momenta
\begin{equation}
  \begin{aligned}
    \hat p^A_1(x) = & +\bar H_{\check 2} -H_{\hat 1} -\bar H_{\check 1}
    \\ &
    +x\frac{\frac{\chi}{\kappa}(G_{\hat 2}(0) -G_{\check 2}(0))+G_{\hat 2}'(0) +G_{\check 2}'(0)}{(x-s)(x+s^{-1})}
    -\frac{1}{2} \frac{\frac{4\pi}{\kappa}\mathcal{L} x}{(x-s)(x+s^{-1})},
\\
    \hat p^S_1(x) = & +H_{\hat 2} - H_{\hat 1} - \bar H_{\check 1}
    \\ &
    -\frac{G_{\hat 2}(0) +G_{\check 2}(0)}{(x-s)(x+s^{-1})}\left(1+\frac{\chi}{\kappa}x\right)-\frac{1}{2} \frac{\frac{4\pi}{\kappa}\mathcal{L} x}{(x-s)(x+s^{-1})},
\\
    \hat p^S_2(x) = & -H_{\hat 2} + H_{\hat 3} +\bar H_{\check 3}
    \\ &
    +\frac{G_{\hat 2}(0) +G_{\check 2}(0)}{(x-s)(x+s^{-1})}\left(1+\frac{\chi}{\kappa}x\right)+\frac{1}{2} \frac{\frac{4\pi}{\kappa}\mathcal{L} x}{(x-s)(x+s^{-1})},
\\
    \hat p^A_2(x) = & -\bar H_{\check 2} + H_{\hat 3} + \bar H_{\check 3}
    \\ &
    - x\frac{\frac{\chi}{\kappa}(G_{\hat 2}(0) -G_{\check 2}(0))+G_{\hat 2}'(0) +G_{\check 2}'(0)}{(x-s)(x+s^{-1})} +\frac{1}{2} \frac{\frac{4\pi}{\kappa}\mathcal{L} x}{(x-s)(x+s^{-1})},
\\
    \check p^S_2(x) = & -\bar H_{\hat 2} + H_{\check 3} + \bar H_{\hat 3}
    \\ &
    -\frac{G_{\hat 2}(0)+G_{\check 2}(0)}{(x+s)(x-s^{-1})}\left(1-\frac{\chi}{\kappa}x\right)
    -\frac{1}{2} \frac{\frac{4\pi}{\kappa}\mathcal{L} x}{(x+s)(x-s^{-1})},
\\
    \check p^A_2(x) = & -H_{\check 2} + H_{\check 3} + \bar H_{\hat 3}
    \\ &
    +x\frac{\frac{\chi}{\kappa}(G_{\hat 2}(0)-G_{\check 2}(0)) +(G_{\hat 2}'(0)+G_{\check 2}'(0))}{(x+s)(x-s^{-1})}-\frac{1}{2} \frac{\frac{4\pi}{\kappa}\mathcal{L} x}{(x+s)(x-s^{-1})},
\\
    \check p^A_1(x) = & +H_{\check 2} - H_{\check 1} - \bar H_{\hat 1}
    \\ &
    -x\frac{\frac{\chi}{\kappa}(G_{\hat 2}(0)-G_{\check 2}(0)) +(G_{\hat 2}'(0)+G_{\check 2}'(0))}{(x+s)(x-s^{-1})}+\frac{1}{2} \frac{\frac{4\pi}{\kappa}\mathcal{L} x}{(x+s)(x-s^{-1})},
\\
    \check p^S_1(x) = & +\bar H_{\hat 2} - H_{\check 1} - \bar H_{\hat 1}
    \\ &
    +\frac{G_{\hat 2}(0)+G_{\check 2}(0)}{(x+s)(x-s^{-1})}\left(1-\frac{\chi}{\kappa}x\right)
    +\frac{1}{2} \frac{\frac{4\pi}{\kappa}\mathcal{L} x}{(x+s)(x-s^{-1})},
  \end{aligned}
\end{equation}
the six finite-gap equations are given by
\begin{equation}
  \begin{aligned}
    \hat p^A_1 - \hat p^S_1 & = 2\pi n_{\hat 1}, \qquad &
    \check p^S_2 - \check p^A_2 &= 2\pi n_{\check 3}, \\
    \hat p^S_1 - \hat p^S_2 & = 2\pi n_{\hat 2}, &
    \check p^A_2 - \check p^A_1 &= 2\pi n_{\check 2},\\
    \hat p^S_2 - \hat p^A_2 & = 2\pi n_{\hat 3}, &
    \check p^A_1 - \check p^S_1 &= 2\pi n_{\check 1}.
  \end{aligned}
\end{equation}
The superscripts $A$ and $S$ indicate whether the quasi-momentum components is related to the AdS or to the sphere's subspace respectively.
The quasi-momenta satisfy the relations
\begin{equation}
  \begin{aligned}
    \hat p^A_1(1/x) &= \check p^A_1(x),\quad &
    \hat p^S_1(1/x) = \check p^S_1(x) + G_{\hat 2}(0)+G_{\check 2}(0) , \\
    \hat p^A_2(1/x) &= \check p^A_2(x),\quad &
    \hat p^S_2(1/x) = \check p^S_2(x) - G_{\hat 2}(0)-G_{\check 2}(0).
  \end{aligned}
\end{equation}
The large $x$ asymptotics gives
\begin{equation}\label{eq:charges}
  \begin{aligned}
    \hat p^A_1(x) & \simeq \frac{1}{\kappa h x} \Bigl(-\frac{1}{2}(D+S) - \frac{1}{2} \hat{B} \Bigr) , & \qquad
    \check p^S_2(x) & \simeq \frac{1}{\kappa h x} \Bigl(-\frac{1}{2}(J-K) - \frac{1}{2} \check{B} \Bigr) ,
    \\
    \hat p^S_1(x) & \simeq \frac{1}{\kappa h x} \Bigl(-\frac{1}{2}(J+K) - \frac{1}{2} \hat{B} \Bigr) , & \qquad
    \check p^A_2(x) & \simeq \frac{1}{\kappa h x} \Bigl(-\frac{1}{2}(D-S) - \frac{1}{2} \check{B} \Bigr) ,
    \\
    \hat p^S_2(x) & \simeq \frac{1}{\kappa h x} \Bigl(+\frac{1}{2}(J+K) - \frac{1}{2} \hat{B} \Bigr) , & \qquad
    \check p^A_1(x) & \simeq \frac{1}{\kappa h x} \Bigl(+\frac{1}{2}(D-S) - \frac{1}{2} \check{B} \Bigr) ,
    \\
    \hat p^A_2(x) & \simeq \frac{1}{\kappa h x} \Bigl(+\frac{1}{2}(D+S) - \frac{1}{2} \hat{B} \Bigr) , & \qquad
    \check p^S_1(x) & \simeq \frac{1}{\kappa h x} \Bigl(+\frac{1}{2}(J-K) - \frac{1}{2} \check{B} \Bigr) .
  \end{aligned}
\end{equation}
The charges $\hat{B} = K_{\hat{1}} - K_{\hat{3}}$ and $\check{B} = K_{\check{1}} - K_{\check{3}}$ are central in $\algPSU(1,1|2)^2$ and hence unphysical.

\begin{figure}
    \newcommand{\FillSheetUpper}[1]{%
      \draw [fill=white,fill opacity=0.75] (+1cm,#1) ++(+4.5cm,0) -- ++(+0.375cm,+0.1875cm) -- ++(-4.5cm,0) -- ++(-0.375cm,-0.1875cm);
    }

    \newcommand{\FillSheetLower}[1]{%
      \draw [fill=white,fill opacity=0.75] (+1cm,#1) -- ++(-0.375cm,-0.1875cm) -- ++(+4.5cm,0) -- ++(+0.375cm,+0.1875cm);
    }

  \centering  
  \begin{tikzpicture}
    [
    node/.style={shape=circle,draw,thick,inner sep=0pt,minimum size=4mm},
    boson/.style={very thick,blue,decoration={snake,amplitude=0.2mm,segment length=1.2mm},decorate},
    fermion/.style={very thick,red,decoration={snake,amplitude=0.2mm,segment length=1.2mm},decorate}
    ]


    \begin{scope}[xshift=-4cm]
      \node (v1) at ( 0cm, +1cm) [node] {};
      \node (v2) at ( 0cm,  0cm) [node] {};
      \node (v3) at ( 0cm, -1cm) [node] {};

      \draw [thick] (v1.south west) -- (v1.north east);
      \draw [thick] (v1.north west) -- (v1.south east);

      \draw [thick] (v3.south west) -- (v3.north east);
      \draw [thick] (v3.north west) -- (v3.south east);

      \draw [thick] (v1) -- (v2);
      \draw [thick] (v2) -- (v3);

      \node at (v2.west) [anchor=east] {\small $+1$};

      \FillSheetUpper{+1.5cm}
      \FillSheetUpper{+0.5cm}
      \FillSheetUpper{-0.5cm}
      \FillSheetUpper{-1.5cm}

      \draw [thick,dashed] (0.625cm,+1cm) -- (+5.4cm,+1cm);
      \draw [thick,dashed] (0.625cm,-1cm) -- (+5.4cm,-1cm);

      \draw [boson] (+2.0cm,+0.5cm) -- (+2.0cm,-0.5cm);
      \draw [boson] (+2.5cm,+1.5cm) -- (+2.5cm,-1.5cm);

      \draw [fermion] (+4.0cm,+1.5cm) -- (+4.0cm,-0.5cm);
      \draw [fermion] (+4.5cm,+0.5cm) -- (+4.5cm,-1.5cm);

      \FillSheetLower{+1.5cm}
      \FillSheetLower{+0.5cm}
      \FillSheetLower{-0.5cm}
      \FillSheetLower{-1.5cm}

      \fill [black] (+2.0cm,+0.5cm) ellipse [x radius=0.05cm,y radius = 0.025cm];
      \fill [black] (+2.0cm,-0.5cm) ellipse [x radius=0.05cm,y radius = 0.025cm];

      \fill [black] (+2.5cm,+1.5cm) ellipse [x radius=0.05cm,y radius = 0.025cm];
      \fill [black] (+2.5cm,-1.5cm) ellipse [x radius=0.05cm,y radius = 0.025cm];

      \fill [black] (+4.0cm,+1.5cm) ellipse [x radius=0.05cm,y radius = 0.025cm];
      \fill [black] (+4.0cm,-0.5cm) ellipse [x radius=0.05cm,y radius = 0.025cm];

      \fill [black] (+4.5cm,+0.5cm) ellipse [x radius=0.05cm,y radius = 0.025cm];
      \draw [black] (+4.5cm,-1.5cm) ellipse [x radius=0.05cm,y radius = 0.025cm];

      \node at (+5.75cm,+1.6cm) [anchor=west] {$\hat{p}_1^A$};
      \node at (+5.75cm,+0.6cm) [anchor=west] {$\hat{p}_1^S$};
      \node at (+5.75cm,-0.6cm) [anchor=west] {$\hat{p}_2^S$};
      \node at (+5.75cm,-1.6cm) [anchor=west] {$\hat{p}_2^A$};
    \end{scope}

    \begin{scope}[xshift=+4cm]
      \node (v1) at ( 0cm, +1cm) [node] {};
      \node (v2) at ( 0cm,  0cm) [node] {};
      \node (v3) at ( 0cm, -1cm) [node] {};

      \draw [thick] (v1.south west) -- (v1.north east);
      \draw [thick] (v1.north west) -- (v1.south east);

      \draw [thick] (v3.south west) -- (v3.north east);
      \draw [thick] (v3.north west) -- (v3.south east);

      \draw [thick] (v1) -- (v2);
      \draw [thick] (v2) -- (v3);

      \node at (v2.west) [anchor=east] {\small $-1$};

      \FillSheetUpper{+1.5cm}
      \FillSheetUpper{+0.5cm}
      \FillSheetUpper{-0.5cm}
      \FillSheetUpper{-1.5cm}

      \draw [thick,dashed] (0.625cm,+1cm) -- (+5.4cm,+1cm);
      \draw [thick,dashed] (0.625cm,-1cm) -- (+5.4cm,-1cm);

      \draw [boson] (+2.0cm,+0.5cm) -- (+2.0cm,-0.5cm);
      \draw [boson] (+2.5cm,+1.5cm) -- (+2.5cm,-1.5cm);

      \draw [fermion] (+4.0cm,+1.5cm) -- (+4.0cm,-0.5cm);
      \draw [fermion] (+4.5cm,+0.5cm) -- (+4.5cm,-1.5cm);

      \FillSheetLower{+1.5cm}
      \FillSheetLower{+0.5cm}
      \FillSheetLower{-0.5cm}
      \FillSheetLower{-1.5cm}

      \fill [black] (+2.0cm,+0.5cm) ellipse [x radius=0.05cm,y radius = 0.025cm];
      \fill [black] (+2.0cm,-0.5cm) ellipse [x radius=0.05cm,y radius = 0.025cm];

      \fill [black] (+2.5cm,+1.5cm) ellipse [x radius=0.05cm,y radius = 0.025cm];
      \fill [black] (+2.5cm,-1.5cm) ellipse [x radius=0.05cm,y radius = 0.025cm];

      \fill [black] (+4.0cm,+1.5cm) ellipse [x radius=0.05cm,y radius = 0.025cm];
      \fill [black] (+4.0cm,-0.5cm) ellipse [x radius=0.05cm,y radius = 0.025cm];

      \fill [black] (+4.5cm,+0.5cm) ellipse [x radius=0.05cm,y radius = 0.025cm];
      \draw [black] (+4.5cm,-1.5cm) ellipse [x radius=0.05cm,y radius = 0.025cm];

      \node at (+5.75cm,+1.6cm) [anchor=west] {$\hat{p}_2^S$};
      \node at (+5.75cm,+0.6cm) [anchor=west] {$\hat{p}_2^A$};
      \node at (+5.75cm,-0.6cm) [anchor=west] {$\hat{p}_1^A$};
      \node at (+5.75cm,-1.6cm) [anchor=west] {$\hat{p}_1^S$};
    \end{scope}
  \end{tikzpicture}

  \caption{\label{fig:sheets}%
    The quasi-momenta form two sets of four-sheeted Riemann surfaces, corresponding to the hatted and checked quasi-momenta. 
    Classical solutions correspond to cuts connecting the different sheets (of the same type, namely hatted or checked).
    The physical excitations correspond to poles connecting the sheets which contain the middle node root.
    Totally there are eight bosonic and fermionic physical excitations with different polarizations which are depicted in the figure by blue and red wiggly lines respectively.
    The lines contain different Dynkin nodes which correspond to the excited Bethe roots.
    The two Riemann surfaces are related by the inversion $x\to 1/x$ symmetry. 
    The $\pm 1$ to the left of the Dynkin diagrams correspond to the grading of the superalgebra, these are the $su(2)$ (left digram) and $sl(2)$ (right diagram) gradings. 
    The dashed lines separate the AdS and sphere's sheets.}

\end{figure}
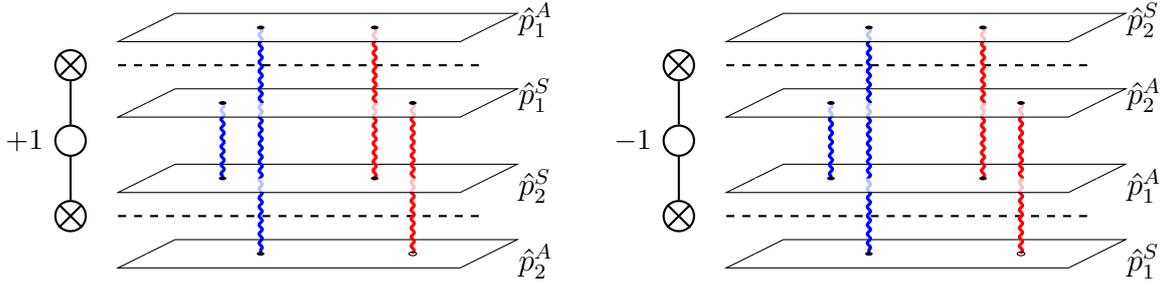

\section{Semi-classical analysis}
\label{sec:semi-classical}

The spectrum of small fluctuations around a classical solution can be found by adding poles to the quasi-momenta, in a way which is consistent with the analytical properties of the quasi-momenta~\cite{Gromov:2007aq}.
Next we adapt the procedure for our case.
The perturb quasi-momenta is given by $p_i(x) + \delta p_i(x)$, where $p_i(x) $ is the quasi-momenta corresponding to a classical string solutions.
The new quasi-momenta will have new microscopic cuts (\ie poles) and macroscopic cuts of poles condensation (related to the classical solution). 
The new poles positions are fixed to leading order by the classical quasi-momenta, however these poles will backreact and shift the macroscopic cuts.
More precisely, along the all the cuts of the Riemann surface $\mathcal{C}_{ij}$ we have
\begin{equation}
(p_i+\delta p_i)^+ - (p_j+\delta p_j)^- = 2\pi n, \quad x\in \mathcal{C}_{ij},
\end{equation}
where the superscript $\pm$ denotes above and below the cut.
This equation fixes the position of the microscopic cuts to leading order by
\begin{equation}\label{eq:polepos}
p_i(x_n^{ij}) - p_j(x_n^{ij})= 2\pi n, \quad |x_n^{ij}|>1,
\end{equation}
where $|x|>1$ is the physical domain, where the quasi-momenta at $|x|<1$ is related by the $\Integers_4$ symmetry.
Then, along the macroscopic cuts the perturbation must satisfy
\begin{equation}
(\delta p_i)^+ - (\delta p_j)^- = 0, \quad x\in \mathcal{C}_n^{ij}.
\end{equation}
As in~\cite{Gromov:2007aq}, we denote the number of excitations with mode number $n$ between the sheets $p_i$ and $p_j$ by $N_n^{ij}$ and $N_{ij} = \sum_n N_n^{ij}$.
The different possible excitations are given by connecting the following sheets (see figure~\ref{fig:sheets})
\begin{align}
\AdS_3:\quad     & (\hat p^A_1,\hat p^A_2), (\check p^A_2,\check p^A_1)\quad \nonumber\\
\Sphere^3:\quad                & (\hat p^S_1,\hat p^S_2), (\check p^S_2,\check p^S_1)\quad \nonumber\\
\text{Fermionic}:\quad   & (\hat p^A_1,\hat p^S_2), (\hat p^S_1,\hat p^A_2), (\check p^A_2,\check p^S_1), (\check p^S_2,\check p^A_1).
\end{align}

The energy shift coming from the poles is given by
\begin{equation}
\delta D = \delta\Delta + \sum_{\mathrm{AdS}_3}N_{ij} + \frac{1}{2}\sum_{\text{fermions}}N_{ij} ,
\end{equation}
where we have extracted the non-trivial piece $\delta\Delta$ from the bare part.
It is then useful to write the asymptotics of the quasi-momenta as
\begin{equation}\label{eq:QMasymp}
  \delta \left(
    \begin{array}{c}
      \hat p_1^A \\
      \hat p_2^A \\[2pt] \hline
      \hat p_1^S \\
      \hat p_2^S \\[2pt] \hline\hline
      \check p_1^A \\
      \check p_2^A \\[2pt] \hline
      \check p_1^S \\
      \check p_2^S \\
    \end{array}
  \right)
  =
  \frac{1}{\kappa h x}\left(
    \begin{array}{l}
      -\frac{1}{2} \delta\Delta - N_{\hat 1 \hat 2}^{A A} - N_{\hat 1 \hat 2}^{A S} \\
      +\frac{1}{2} \delta\Delta + N_{\hat 1 \hat 2}^{A A} + N_{\hat 1 \hat 2}^{S A} \\[2pt] \hline
      \phantom{-\frac{1}{2} \delta\Delta} + N_{\hat 1 \hat 2}^{S S} + N_{\hat 1 \hat 2}^{S A} \\
      \phantom{-\frac{1}{2} \delta\Delta} - N_{\hat 1 \hat 2}^{S S} - N_{\hat 1 \hat 2}^{A S} \\[2pt] \hline\hline
      +\frac{1}{2} \delta\Delta + N_{\check 2 \check 1}^{A A} + N_{\check 2 \check 1}^{A S} \\
      -\frac{1}{2} \delta\Delta - N_{\check 2 \check 1}^{A A} - N_{\check 2 \check 1}^{S A} \\[2pt] \hline
      \phantom{-\frac{1}{2} \delta\Delta} - N_{\check 2 \check 1}^{S S} - N_{\check 2 \check 1}^{S A} \\
      \phantom{-\frac{1}{2} \delta\Delta} + N_{\check 2 \check 1}^{S S} + N_{\check 2 \check 1}^{A S}
    \end{array}
  \right) .
\end{equation}
The filling fractions $N_n^{ij}$ are related by
\begin{equation}
  \sum_n n \sum_{i,j} N_n^{ij} = 0.
\end{equation}
The signs of the residues are given by
\begin{equation}
  \begin{aligned}\label{eq:reseq}
    \operatorname*{res}_{x=x_n^{12}} \hat p_i^A &   = -(\delta_{1 i} - \delta_{2 i})\hat \alpha(x_n^{12})N_n^{12}, \\
    \operatorname*{res}_{x=x_n^{12}} \hat p_i^S &   = +(\delta_{1 i} - \delta_{2 i})\hat \alpha(x_n^{12})N_n^{12}, \\
    \operatorname*{res}_{x=x_n^{12}} \check p_i^A & = +(\delta_{1 i} - \delta_{2 i})\check \alpha(x_n^{12})N_n^{12}, \\
    \operatorname*{res}_{x=x_n^{12}} \check p_i^S & = -(\delta_{1 i} - \delta_{2 i})\check \alpha(x_n^{12})N_n^{12}.
  \end{aligned}
\end{equation}
The perturbed quasi-momenta also have to satisfy the $\Integers_4$ inversion symmetry, so that
\begin{equation}
  \begin{aligned}
    \delta \hat p^A_1(1/x) &= \delta \check p^A_1(x),\quad &
    \delta \hat p^S_1(1/x) &= \delta \check p^S_1(x), \\
    \delta \hat p^A_2(1/x) &= \delta \check p^A_2(x),\quad &
    \delta \hat p^S_2(1/x) &= \delta \check p^S_2(x).
  \end{aligned}
\end{equation}
Finally, the poles at $\pm s$ and $\pm 1/s$ are synchronized in the following way
\begin{equation}\label{eq:poles}
  \delta(\hat p_1^A, \hat p_2^A | \hat p_1^S, \hat p_2^S || \check p_1^A, \check p_2^A | \check p_1^S, \check p_2^S)
  \simeq
  \begin{cases}
    +s\frac{(\delta\alpha_{+}, \delta\beta_{+} | \delta\alpha_{+}, \delta\beta_{+} ||0,0,0,0)}{x-s} \\
    -\frac{1}{s}\frac{(\delta\alpha_{-}, \delta\beta_{-} | \delta\alpha_{-}, \delta\beta_{-} ||0,0,0,0)}{x+1/s} \\
    +s\frac{(0,0,0,0 || \delta\alpha_{-}, \delta\beta_{-} | \delta\alpha_{-}, \delta\beta_{-})}{x+s} \\
    -\frac{1}{s}\frac{(0,0,0,0 || \delta\alpha_{+}, \delta\beta_{+} | \delta\alpha_{+}, \delta\beta_{+})}{x-1/s} \\
  \end{cases}
\end{equation}
where we have used~\eqref{eq:residues} and~\eqref{eq:MConeformsgrading}.

\subsection{BMN string quantization}
\label{sec:BMN-quantization}

Next we apply the above procedure to the simplest case of the BMN string solution where the classical solution has no cuts.
The analysis is similar to the one in~\cite{Gromov:2007aq}.
In this section we will give a sketch of the procedure as well as the result for the energy fluctuations, while saving the details for appendix~\ref{app:BMNquantization}.
The classical quasi-momenta is given by
\begin{equation}
  p_l(x) = (p(x),-p(x)|p(x),-p(x)||p(1/x),-p(1/x),p(1/x),-p(1/x))
\end{equation}
with
\begin{equation}
  p(x) = \frac{2\pi x \mathcal{J}}{\kappa(x-s)(x-s^{-1})},
\end{equation}
where $\mathcal{J}$ is the angular momentum of the BMN ground state.
Using~\eqref{eq:polepos} we fix the position of the poles
\begin{equation}
  x_n^{\hat \imath\hat \jmath} = \frac{
    \mathcal{J} + \chi n + \sqrt{
      \mathcal{J}^2 + 2 \chi \mathcal{J} n + n^2
    }}{ \kappa n }
\end{equation}
for the hatted quasi-momenta, which correspond to the AdS excitations, and similarly to the sphere's and fermionic excitations.
Next we make an ansatz for the perturbed quasi-momenta, such that it has the desired pole structure according to~\eqref{eq:reseq} and~\eqref{eq:poles} and satisfies the $\Integers_4$ inversion symmetry.
The only unknowns are the residues at poles~\eqref{eq:poles} which however are synchronized, and an overall additive constant to the quasi-momenta.
These unknowns are later fixed by using the large $x$ asymptotics~\eqref{eq:QMasymp}, and yield the energy fluctuations to be 
\begin{equation}
  \delta \Delta  =
  \sum_{\text{all}~ij}\sum_n
  \left(
    \hat N^n_{ij} \left(
      \sqrt{ \frac{n^2}{\mathcal{J}^2} + 2 \chi \frac{n}{\mathcal{J}} +1} - 1
    \right)
    +
    \check N^n_{ij} \left(
      \sqrt{ \frac{n^2}{\mathcal{J}^2} - 2 \chi \frac{n}{\mathcal{J}} +1} - 1
    \right)\right).
\end{equation}
This result is consistent with previous results for quantization of the BMN string with mixed flux~\cite{Berenstein:2002jq}.

\section{One-loop corrections to the dressing phase}
\label{sec:phase}

As discussed in section~\ref{sec:finite-gap-from-BA}, the two-particle S-matrix contain two undetermined scalar factors $\sigma$ and $\bar{\sigma}$. In that section we also gave a proposal for the leading order expressions for these phases. Here we will calculate the one-loop correction to the phases by quantizing the algebraic curve around a generic classical solution. This method was first used in the $\AdS_5 \times \Sphere^5$ setting in~\cite{Gromov:2007cd}, and has also been used to calculate the one-loop phases in $\AdS_3 \times \Sphere^3 \times \Torus^4$ with only RR flux~\cite{Beccaria:2012kb}.

To see how the dressing phases enter in the algebraic curve calculation, let us consider the Bethe equations~\eqref{eq:BA-2a} for the Bethe roots $\hat{x}_{2,k}^{\pm}$. The phase factors enter in the combination
\begin{equation}
  \prod_{\substack{j = 1 \\ j \neq k}}^{\hat{K}_2} \sigma^2(\hat{x}^\pm_{2mk},\hat{x}^\pm_{2,j})
  \prod_{j=1}^{\check{K}_2} \bar{\sigma}^2(\hat{x}^\pm_{2,k},\check{x}^\pm_{2,j}) .
\end{equation}
It is useful to express $\sigma$ and $\bar{\sigma}$ in terms of actual phases by introducing two functions $\theta$ and $\bar{\theta}$ by
\begin{equation}
  \sigma(\hat{x}^\pm,\hat{y}^\pm) = \exp \bigl( i \theta(\hat{x}^\pm,\hat{y}^\pm) \bigr) ,
  \qquad
  \bar{\sigma}(\hat{x}^\pm,\check{y}^\pm) = \exp \bigl( i \bar{\theta}(\hat{x}^\pm,\check{y}^\pm) \bigr) .
\end{equation}
The phases $\theta$ and $\bar{\theta}$ are functions of the coupling constant $h$ and can be expanded at large coupling
\begin{equation}
  \begin{aligned}
    \theta(\hat{x}^\pm,\hat{y}^\pm) &= h \, \theta^{(0)}(\hat{x}^\pm,\hat{y}^\pm) + \theta^{(1)}(\hat{x}^\pm,\hat{y}^\pm) + \mathcal{O}(1/h) ,
    \\ 
    \bar{\theta}(\hat{x}^\pm,\check{y}^\pm) &= h \, \bar{\theta}^{(0)}(\hat{x}^\pm,\check{y}^\pm) + \bar{\theta}^{(1)}(\hat{x}^\pm,\check{y}^\pm) + \mathcal{O}(1/h) ,
  \end{aligned}
\end{equation}
where the superscript indicates the order in the large $h$ expansion. The leading terms $\theta^{(0)}$ and $\bar{\theta}^{(0)}$ were discussed in section~\ref{sec:Bethe-ansatz} and are already included in the finite-gap equations. To also include the one-loop phase we can add potentials $-2\hat{\mathcal{V}}(x)$ and $-2\check{\mathcal{V}}(x)$ to the right hand sides of equations~\eqref{eq:finite-gap-2a} and~\eqref{eq:finite-gap-2b}, with
\begin{equation}
  \begin{aligned}
    \hat{\mathcal{V}}(\hat{x}) &= \sum_{j=1}^{\hat{K}_2} \theta^{(1)}(\hat{x},\hat{x}_{2,j}) + \sum_{j=1}^{\check{K}_2} \bar{\theta}^{(1)}(\hat{x},\check{x}_{2,j}) , \\
    \check{\mathcal{V}}(\check{x}) &= \sum_{j=1}^{\check{K}_2} \theta^{(1)}(\check{x},\check{x}_{2,j}) + \sum_{j=1}^{\hat{K}_2} \bar{\theta}^{(1)}(\check{x},\hat{x}_{2,j}).
  \end{aligned}
\end{equation}
By expressing the finite-gap equations in terms of quasi-momenta $\hat{p}_i$ and $\check{p}_i$ we see that the result of the one-loop phases is to shift the quasi-momenta by plus or minus the above potentials. These shift originates in the leading quantum corrections to the string state. Following~\cite{Gromov:2007cd}, we find the one-loop correction to the dressing phase by first considering a more general setting where we add different potentials $\hat{\mathcal{V}}_i$ and $\check{\mathcal{V}}_i$ to each quasi-momentum. These potentials are obtained from the one-loop shift of the energy of a generic string state, which we get as a graded sum over all fluctuations. It is useful to distinguish the contribution to the potentials coming from hatted and checked excitations. If we add a pole at position $x_n^{\hat{\imath}\hat{\jmath}}$ between the sheets $\hat{\imath}$ and $\hat{\jmath}$ we get
\begin{equation}
  \begin{aligned}
    \begin{pmatrix}
      \hat{\mathcal{V}}_1^A \\ \hat{\mathcal{V}}_1^S \\ \hat{\mathcal{V}}_2^S \\ \hat{\mathcal{V}}_2^A
    \end{pmatrix}^{\hat{\imath}\hat{\jmath}} \!
    &=
    \frac{\hat{\alpha}(x)}{x^2}
    \begin{pmatrix}
      - x \hat{d} \\ + \bigl( 1 + \frac{\chi}{\kappa} x \bigr) \hat{c} \\ - \bigl( 1 + \frac{\chi}{\kappa} x \bigr) \hat{c} \\ + x \hat{d}
    \end{pmatrix}
    +
    \frac{\hat{\alpha}(x)}{x - x_n^{\hat{\imath}\hat{\jmath}}}
    \begin{pmatrix}
      - \delta_{\hat{\imath},\hat{1}^A} \\ + \delta_{\hat{\imath},\hat{1}^S} \\ - \delta_{\hat{\jmath},\hat{2}^S} \\ + \delta_{\hat{\jmath},\hat{2}^A}
    \end{pmatrix} ,
    \\
    \begin{pmatrix}
      \check{\mathcal{V}}_2^S \\ \check{\mathcal{V}}_2^A \\ \check{\mathcal{V}}_1^A \\ \check{\mathcal{V}}_1^S
    \end{pmatrix}^{\hat{\imath}\hat{\jmath}} \!
    &=
    \frac{\check{\alpha}(x)}{x^2}
    \begin{pmatrix}
      + \bigl( 1 - \frac{\chi}{\kappa} x \bigr) \hat{c} \\ - x \hat{d} \\ + x \hat{d} \\ - \bigl( 1 - \frac{\chi}{\kappa} x \bigr) \hat{c}
    \end{pmatrix}
    +
    \frac{\hat{\alpha}(1/x)}{\frac{1}{x} - x_n^{\hat{\imath}\hat{\jmath}}}
    \begin{pmatrix}
      - \delta_{\hat{\jmath},\hat{2}^S} \\ + \delta_{\hat{\jmath},\hat{2}^A} \\ - \delta_{\hat{\imath},\hat{1}^A} \\ + \delta_{\hat{\imath},\hat{1}^S}
    \end{pmatrix} ,
  \end{aligned}
\end{equation}
where
\begin{equation}
  \hat{d} = \frac{\hat{\alpha}(x_n^{\hat{\imath}\hat{\jmath}})}{(x_n^{\hat{\imath}\hat{\jmath}})^2} \bigl( 1 + \frac{\chi}{\kappa} x_n^{\hat{\imath}\hat{\jmath}} \bigr) , \qquad
  \hat{c} = \frac{\hat{\alpha}(x_n^{\hat{\imath}\hat{\jmath}})}{x_n^{\hat{\imath}\hat{\jmath}}} .
\end{equation}
Similarly an additional pole between sheets $\check{\imath}$ and $\check{\jmath}$ gives
\begin{equation}
  \begin{aligned}
    \begin{pmatrix}
      \check{\mathcal{V}}_2^S \\ \check{\mathcal{V}}_2^A \\ \check{\mathcal{V}}_1^A \\ \check{\mathcal{V}}_1^S
    \end{pmatrix}^{\check{\imath}\check{\jmath}} \!
    &=
    \frac{\check{\alpha}(x)}{x^2}
    \begin{pmatrix}
      + \bigl( 1 - \frac{\chi}{\kappa} x \bigr) \check{c} \\ - x \check{d} \\ + x \check{d} \\ - \bigl( 1 - \frac{\chi}{\kappa} x \bigr) \check{c}
    \end{pmatrix}
    +
    \frac{\check{\alpha}(x)}{x - x_n^{\check{\imath}\check{\jmath}}}
    \begin{pmatrix}
      + \delta_{\check{\imath},\check{2}^S} \\ - \delta_{\check{\imath},\check{2}^A} \\ + \delta_{\check{\jmath},\check{1}^A} \\ - \delta_{\check{\jmath},\check{1}^S}
    \end{pmatrix} ,
    \\
    \begin{pmatrix}
      \hat{\mathcal{V}}_1^A \\ \hat{\mathcal{V}}_1^S \\ \hat{\mathcal{V}}_2^S \\ \hat{\mathcal{V}}_2^A
    \end{pmatrix}^{\check{\imath}\check{\jmath}} \!
    &=
    \frac{\hat{\alpha}(x)}{x^2}
    \begin{pmatrix}
      - x \check{d} \\ + \bigl( 1 + \frac{\chi}{\kappa} x \bigr) \check{c} \\ - \bigl( 1 + \frac{\chi}{\kappa} x \bigr) \check{c} \\ + x \check{d}
    \end{pmatrix}
    +
    \frac{\check{\alpha}(1/x)}{\frac{1}{x} - x_n^{\check{\imath}\check{\jmath}}}
    \begin{pmatrix}
      + \delta_{\check{\jmath},\check{1}^A} \\ - \delta_{\check{\jmath},\check{1}^S} \\ + \delta_{\check{\imath},\check{2}^S} \\ - \delta_{\check{\imath},\check{2}^A}
    \end{pmatrix} ,
  \end{aligned}
\end{equation}
with
\begin{equation}
  \check{d} = \frac{\check{\alpha}(x_n^{\check{\imath}\check{\jmath}})}{(x_n^{\check{\imath}\check{\jmath}})^2} \bigl( 1 - \frac{\chi}{\kappa} x_n^{\check{\imath}\check{\jmath}} \bigr) , \qquad
  \check{c} = \frac{\check{\alpha}(x_n^{\check{\imath}\check{\jmath}})}{x_n^{\check{\imath}\check{\jmath}}} .
\end{equation}
The full potentials are then given by a graded sum over all fluctuations. Let us consider
\begin{equation}
  \hat{\mathcal{V}}_k = \frac{1}{2} \sum_{n=-\infty}^{+\infty} \Bigl(
  \sum_{\hat{\imath}\hat{\jmath}} (-1)^F \hat{\mathcal{V}}_k^{\hat{\imath}\hat{\jmath}}
  + \sum_{\check{\imath}\check{\jmath}} (-1)^F \hat{\mathcal{V}}_k^{\check{\imath}\check{\jmath}}
  \Bigr) .
\end{equation}
We can rewrite the sums over $n$ as the integrals
\begin{equation}
  \hat{\mathcal{V}}_k =
  \frac{1}{4i} \int_{\hat{\mathcal{C}}} dn \cot(\pi n) \Bigl( \sum_{\hat{\imath}\hat{\jmath}} (-1)^F \hat{\mathcal{V}}_k^{\hat{\imath}\hat{\jmath}} \Bigr)
  + \frac{1}{4i} \int_{\check{\mathcal{C}}} dn \cot(\pi n) \Bigl( \sum_{\check{\imath}\check{\jmath}} (-1)^F \hat{\mathcal{V}}_k^{\check{\imath}\check{\jmath}} \Bigr)
  \Bigr) ,
\end{equation}
where the integration contours $\hat{\mathcal{C}}$ and $\check{\mathcal{C}}$ encircle all the poles of $\cot$. To understand these contours let us consider the BMN frequencies from the previous section,
\begin{equation}
  \hat{\mathcal{E}}_n = \sqrt{\left(\frac{n}{\mathcal{J}}\right)^2+2 \chi \frac{n}{\mathcal{J}} +1} - 1 , \qquad
  \check{\mathcal{E}}_n = \sqrt{\left(\frac{n}{\mathcal{J}}\right)^2-2 \chi \frac{n}{\mathcal{J}} +1} - 1 .
\end{equation}
\begin{figure}
  \centering
  \subfloat[\label{fig:int-contour-n}]{

  \begin{tikzpicture}[
    scale=2,
    cut/.style={thick,blue,decorate,decoration={snake,segment length=1mm,amplitude=.2mm}},
    contour/.style={thick,red},
    cross/.style={thick,line cap=round}
    ]

    \useasboundingbox [clip] (-1.55,-1.45) rectangle (1.55,1.45);

    \draw [->] (-1.5,0) -- (+1.5,0);
    \draw [->] (0,-1.4) -- (0,+1.4);

    \draw [cut] (-120:1) +(0,-0.8) -- +(0,+0.4);
    \draw [cut] (+120:1) +(0,+0.8) -- +(0,-0.4);

    \draw [cross,blue] (+120:1) ++(0,-0.4) +(+135:0.04) -- +(-45:0.04) +(-135:0.04) -- +(+45:0.04);
    \draw [cross,blue] (-120:1) ++(0,+0.4) +(+135:0.04) -- +(-45:0.04) +(-135:0.04) -- +(+45:0.04);

    \draw [cross,violet] (-1.33,0) +(+135:0.04) -- +(-45:0.04) +(-135:0.04) -- +(+45:0.04);
    \draw [cross,violet] (-1.11,0) +(+135:0.04) -- +(-45:0.04) +(-135:0.04) -- +(+45:0.04);
    \draw [cross,violet] (-0.88,0) +(+135:0.04) -- +(-45:0.04) +(-135:0.04) -- +(+45:0.04);
    \draw [cross,violet] (-0.66,0) +(+135:0.04) -- +(-45:0.04) +(-135:0.04) -- +(+45:0.04);
    \draw [cross,violet] (-0.44,0) +(+135:0.04) -- +(-45:0.04) +(-135:0.04) -- +(+45:0.04);
    \draw [cross,violet] (-0.22,0) +(+135:0.04) -- +(-45:0.04) +(-135:0.04) -- +(+45:0.04);
    \draw [cross,violet] ( 0.00,0) +(+135:0.04) -- +(-45:0.04) +(-135:0.04) -- +(+45:0.04);
    \draw [cross,violet] (+0.22,0) +(+135:0.04) -- +(-45:0.04) +(-135:0.04) -- +(+45:0.04);
    \draw [cross,violet] (+0.44,0) +(+135:0.04) -- +(-45:0.04) +(-135:0.04) -- +(+45:0.04);
    \draw [cross,violet] (+0.66,0) +(+135:0.04) -- +(-45:0.04) +(-135:0.04) -- +(+45:0.04);
    \draw [cross,violet] (+0.88,0) +(+135:0.04) -- +(-45:0.04) +(-135:0.04) -- +(+45:0.04);
    \draw [cross,violet] (+1.11,0) +(+135:0.04) -- +(-45:0.04) +(-135:0.04) -- +(+45:0.04);
    \draw [cross,violet] (+1.33,0) +(+135:0.04) -- +(-45:0.04) +(-135:0.04) -- +(+45:0.04);

    \draw (+120:1) +(0.05,-0.415) node [anchor=west] {$\scriptstyle n_+$};
    \draw (-120:1) +(0.05,+0.385) node [anchor=west] {$\scriptstyle n_-$};

    \draw [contour] (-1,0) arc (+180:+124:1) -- ++(0,-0.4) arc (-180:0:0.06) -- (+116:1) arc (+116:0:1);
    \draw [contour] (-1,0) arc (-180:-124:1) -- ++(0,+0.4) arc (+180:0:0.06) -- (-116:1) arc (-116:0:1);

    \draw [contour,->] ([shift=(+ 40:1)] 0,0) arc (+ 40:+ 45:1);
    \draw [contour,->] ([shift=(+ 85:1)] 0,0) arc (+ 85:+ 90:1);
    \draw [contour,->] ([shift=(+130:1)] 0,0) arc (+130:+135:1);
    \draw [contour,->] ([shift=(+175:1)] 0,0) arc (+175:+180:1);
    \draw [contour,->] ([shift=(- 95:1)] 0,0) arc (- 95:- 90:1);
    \draw [contour,->] ([shift=(-140:1)] 0,0) arc (-140:-135:1);
    \draw [contour,->] ([shift=(- 50:1)] 0,0) arc (- 50:- 45:1);
    \draw [contour,->] ([shift=(-  5:1)] 0,0) arc (-  5:   0:1);

    \node at (-1,0.08) [anchor=west] {$\scriptstyle -N$};
    \node at (+1,0.08) [anchor=east] {$\scriptstyle +N$};

    \draw (1.1,1.1) node [anchor=center] {$n$} ++(-0.09,-0.09) ++ (0,0.18) -- ++(0,-0.18) -- ++(0.18,0);

  \end{tikzpicture}

   }
   \hspace{0.5cm}
   \subfloat[\label{fig:int-contour-x}]{
  \begin{tikzpicture}[
    scale=2,
    cut/.style={thick,blue,decorate,decoration={snake,segment length=1mm,amplitude=.2mm}},
    contour/.style={thick,red},
    cross/.style={thick,line cap=round}
    ]

    \useasboundingbox [clip] (-1.25,-1.45) rectangle (2.25,1.45);

    \draw [->] (-1.15,0) -- (+2.15,0);
    \draw [->] (0,-1.4) -- (0,+1.4);

    \draw [cut] ([shift=(-120:1)] 0.5,0) arc (-120:+120:1);
    \draw [cut] ([shift=(+120:1)] 0.5,0) arc (+120:+240:1);

    \draw [contour] ([shift=(+45:1.1)] 0.5,0) arc (+45:+175:1.1) arc (90:270:0.09587) arc (+185:+355:1.1) arc (-90:+90:0.09587) arc (5:45:1.1);

    \draw [contour,->] ([shift=(- 40:1.1)] 0.5,0) arc (- 40:- 45:1.1);
    \draw [contour,->] ([shift=(- 85:1.1)] 0.5,0) arc (- 85:- 90:1.1);
    \draw [contour,->] ([shift=(-130:1.1)] 0.5,0) arc (-130:-135:1.1);
    \draw [contour,->] ([shift=(+140:1.1)] 0.5,0) arc (+140:+135:1.1);
    \draw [contour,->] ([shift=(+ 95:1.1)] 0.5,0) arc (+ 95:+ 90:1.1);
    \draw [contour,->] ([shift=(+ 50:1.1)] 0.5,0) arc (+ 50:+ 45:1.1);

    \node at (0,+0.78) [anchor=west] {$\scriptstyle +i$};
    \node at (0,-0.78) [anchor=west] {$\scriptstyle -i$};

    \draw [cross,blue] ([shift=(+120:1)] 0.5,0) +(+75:0.04) -- +(-105:0.04) +(+165:0.04) -- +(-15:0.04);
    \draw [cross,blue] ([shift=(-120:1)] 0.5,0) +(+195:0.04) -- +(+15:0.04) +(-75:0.04) -- +(+105:0.04);

    \draw [cross,violet] (-1.05,0) +(+135:0.04) -- +(-45:0.04) +(-135:0.04) -- +(+45:0.04);
    \draw [cross,violet] (-0.85,0) +(+135:0.04) -- +(-45:0.04) +(-135:0.04) -- +(+45:0.04);
    \draw [cross,violet] (-0.74,0) +(+135:0.04) -- +(-45:0.04) +(-135:0.04) -- +(+45:0.04);
    \draw [cross,violet] (-0.64,0) +(+135:0.04) -- +(-45:0.04) +(-135:0.04) -- +(+45:0.04);
    \draw [cross,violet] (-0.59,0) +(+135:0.04) -- +(-45:0.04) +(-135:0.04) -- +(+45:0.04);

    \draw [cross,violet] (+1.59,0) +(+135:0.04) -- +(-45:0.04) +(-135:0.04) -- +(+45:0.04);
    \draw [cross,violet] (+1.64,0) +(+135:0.04) -- +(-45:0.04) +(-135:0.04) -- +(+45:0.04);
    \draw [cross,violet] (+1.74,0) +(+135:0.04) -- +(-45:0.04) +(-135:0.04) -- +(+45:0.04);
    \draw [cross,violet] (+1.85,0) +(+135:0.04) -- +(-45:0.04) +(-135:0.04) -- +(+45:0.04);
    \draw [cross,violet] (+2.05,0) +(+135:0.04) -- +(-45:0.04) +(-135:0.04) -- +(+45:0.04);

    \fill (-0.515,0) circle (0.035);
    \fill (+1.52,0) circle (0.035);

    \fill (+0.5,0) circle (0.015);

    \draw [very thin,decorate,decoration={brace}] (0.5,0) -- (0,0);
    \draw [very thin,decorate,decoration={brace}] (1.5,0) -- (0.5,0);

    \node at (+0.25,0) [anchor=north] {$\scriptstyle \chi/\kappa$};
    \node at (+1,0) [anchor=north] {$\scriptstyle 1/\kappa$};

    \node at (-0.515,0.08) [anchor=west] {$\scriptstyle -1/s$};
    \node at (+1.52,0.08) [anchor=east] {$\scriptstyle +s$};

    \draw (1.6,1.1) node [anchor=center] {$x$} ++(-0.09,-0.09) ++ (0,0.18) -- ++(0,-0.18) -- ++(0.18,0);

   \end{tikzpicture}
  }

  \caption{\label{fig:int-contours} Integration contours in the $n$ and $x$ planes. The blue wavy lines indicate the two square root branch cuts with brach points $n=n_{\pm}$ and $x = \mp i$. The solid red line is the integration contour, which in the $n$ plane is taken along a circle that has been deformed to avoid the branch cuts. This picks up the poles of the $\cot$ function, indicated in the figure by purple crosses along the real axis. For large $N$ the contour in the $x$ plane approaches the branch cuts from the outside.}
\end{figure}
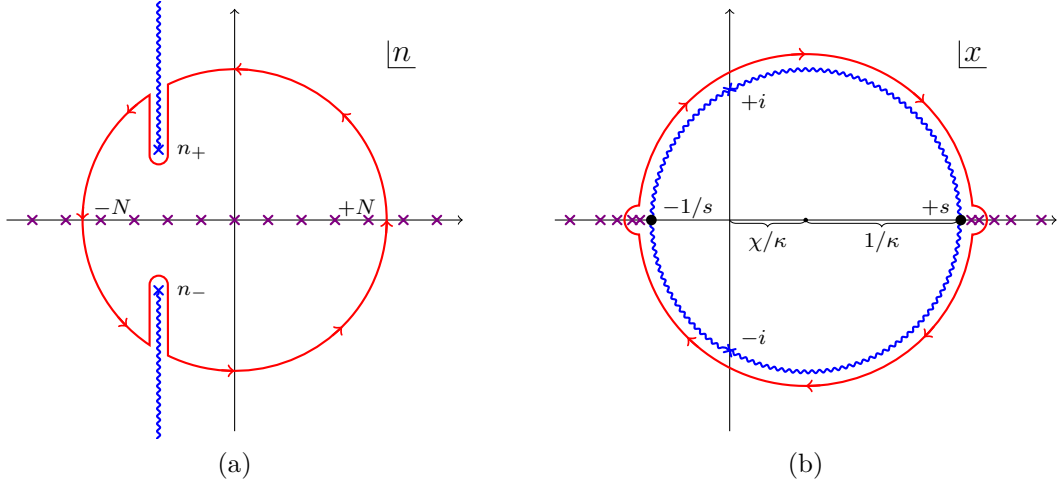%
The function $\hat{\mathcal{E}}_n$ has two branch cuts starting at the points $n_{\pm} = \pm i \mathcal{J} ( \kappa \pm i \chi )$ and running off to infinity. The sum over frequencies from $-N$ to $+N$ can be obtained by integrating over a circle of radius $N$ with two deformations to avoid the branch cuts, see figure~\ref{fig:int-contour-n}. Using the relation
\begin{equation}
  p_i(x_n^{ij}) - p_j(x_n^{ij}) = 2\pi n
\end{equation}
we can map this contour to the $x$ plane, see figure~\ref{fig:int-contour-x}. The branch points $n_{\pm}$ are then mapped to $\mp i$ and the cuts form a circle passing through these points as well as $+s$ and $-1/s$. As $N$ approaches infinity the integration contour $\hat{\mathcal{C}}$ encircles the cuts, avoiding the simple poles at $+s$ and $-1/s$.
For the checked fluctuations frequencies $\check{\mathcal{E}}_n$ we find the same structure as for $\hat{\mathcal{E}}_n$, except we need to replace $\chi \to -\chi$, which shifts the branch cuts in figure~\ref{fig:int-contour-n} from the negative to the positive half plane.

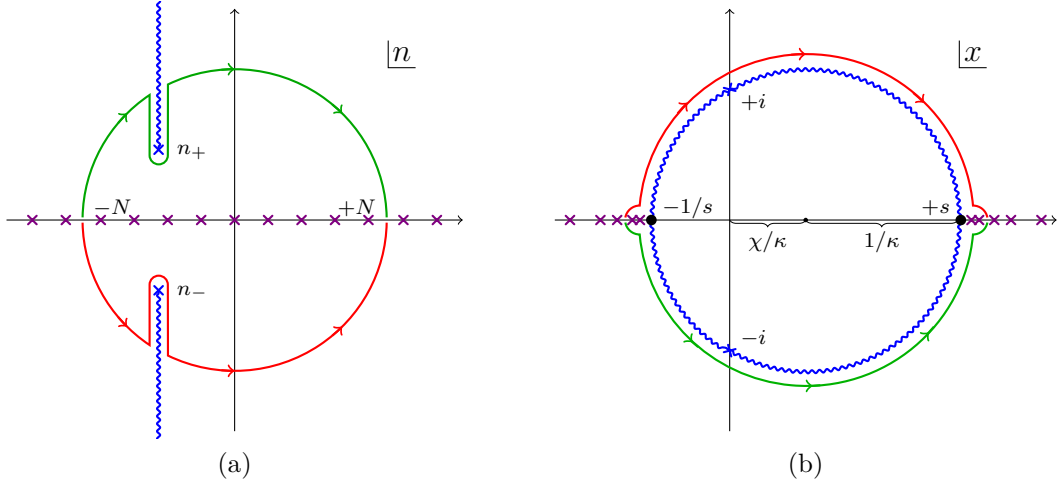
\begin{figure}
  \centering
  
  \subfloat[\label{fig:int-contour-sgn-n}]{

  \begin{tikzpicture}[
    scale=2,
    cut/.style={thick,blue,decorate,decoration={snake,segment length=1mm,amplitude=.2mm}},
    red contour/.style={thick,red},
    green contour/.style={thick,green!65!black},
    cross/.style={thick,line cap=round}
    ]

    \useasboundingbox [clip] (-1.55,-1.45) rectangle (1.55,1.45);

    \draw [->] (-1.5,0) -- (+1.5,0);
    \draw [->] (0,-1.4) -- (0,+1.4);

    \draw [cut] (-120:1) +(0,-0.8) -- +(0,+0.4);
    \draw [cut] (+120:1) +(0,+0.8) -- +(0,-0.4);

    \draw [cross,blue] (+120:1) ++(0,-0.4) +(+135:0.04) -- +(-45:0.04) +(-135:0.04) -- +(+45:0.04);
    \draw [cross,blue] (-120:1) ++(0,+0.4) +(+135:0.04) -- +(-45:0.04) +(-135:0.04) -- +(+45:0.04);

    \draw [cross,violet] (-1.33,0) +(+135:0.04) -- +(-45:0.04) +(-135:0.04) -- +(+45:0.04);
    \draw [cross,violet] (-1.11,0) +(+135:0.04) -- +(-45:0.04) +(-135:0.04) -- +(+45:0.04);
    \draw [cross,violet] (-0.88,0) +(+135:0.04) -- +(-45:0.04) +(-135:0.04) -- +(+45:0.04);
    \draw [cross,violet] (-0.66,0) +(+135:0.04) -- +(-45:0.04) +(-135:0.04) -- +(+45:0.04);
    \draw [cross,violet] (-0.44,0) +(+135:0.04) -- +(-45:0.04) +(-135:0.04) -- +(+45:0.04);
    \draw [cross,violet] (-0.22,0) +(+135:0.04) -- +(-45:0.04) +(-135:0.04) -- +(+45:0.04);
    \draw [cross,violet] ( 0.00,0) +(+135:0.04) -- +(-45:0.04) +(-135:0.04) -- +(+45:0.04);
    \draw [cross,violet] (+0.22,0) +(+135:0.04) -- +(-45:0.04) +(-135:0.04) -- +(+45:0.04);
    \draw [cross,violet] (+0.44,0) +(+135:0.04) -- +(-45:0.04) +(-135:0.04) -- +(+45:0.04);
    \draw [cross,violet] (+0.66,0) +(+135:0.04) -- +(-45:0.04) +(-135:0.04) -- +(+45:0.04);
    \draw [cross,violet] (+0.88,0) +(+135:0.04) -- +(-45:0.04) +(-135:0.04) -- +(+45:0.04);
    \draw [cross,violet] (+1.11,0) +(+135:0.04) -- +(-45:0.04) +(-135:0.04) -- +(+45:0.04);
    \draw [cross,violet] (+1.33,0) +(+135:0.04) -- +(-45:0.04) +(-135:0.04) -- +(+45:0.04);

    \draw (+120:1) +(0.05,-0.415) node [anchor=west] {$\scriptstyle n_+$};
    \draw (-120:1) +(0.05,+0.385) node [anchor=west] {$\scriptstyle n_-$};

    \draw [green contour] (+179:1) arc (+179:+124:1) -- ++(0,-0.4) arc (-180:0:0.06) -- (+116:1) arc (+116:+1:1);
    \draw [red contour]   (-179:1) arc (-179:-124:1) -- ++(0,+0.4) arc (+180:0:0.06) -- (-116:1) arc (-116:-1:1);

    \draw [green contour,->] ([shift=(+ 95:1)] 0,0) arc (+ 95:+ 90:1);
    \draw [green contour,->] ([shift=(+140:1)] 0,0) arc (+140:+135:1);
    \draw [green contour,->] ([shift=(+ 50:1)] 0,0) arc (+ 50:+ 45:1);

    \draw [red contour,->] ([shift=(- 95:1)] 0,0) arc (- 95:- 90:1);
    \draw [red contour,->] ([shift=(-140:1)] 0,0) arc (-140:-135:1);
    \draw [red contour,->] ([shift=(- 50:1)] 0,0) arc (- 50:- 45:1);

    \node at (-1,0.08) [anchor=west] {$\scriptstyle -N$};
    \node at (+1,0.08) [anchor=east] {$\scriptstyle +N$};

    \draw (1.1,1.1) node [anchor=center] {$n$} ++(-0.09,-0.09) ++ (0,0.18) -- ++(0,-0.18) -- ++(0.18,0);

  \end{tikzpicture}

   }
   \hspace{0.5cm}
   \subfloat[\label{fig:int-contour-sgn-x}]{

     \begin{tikzpicture}[
       scale=2,
       cut/.style={thick,blue,decorate,decoration={snake,segment length=1mm,amplitude=.2mm}},
       red contour/.style={thick,red},
       green contour/.style={thick,green!70!black},
       cross/.style={thick,line cap=round}
       ]

       \useasboundingbox [clip] (-1.25,-1.45) rectangle (2.25,1.45);

       \draw [->] (-1.15,0) -- (+2.15,0);
       \draw [->] (0,-1.4) -- (0,+1.4);

       \draw [cut] ([shift=(-120:1)] 0.5,0) arc (-120:+120:1);
       \draw [cut] ([shift=(+120:1)] 0.5,0) arc (+120:+240:1);


       \draw [red contour]   (1.6,0) ++ (+10:0.09587) arc (+10:+90:0.09587) arc (+5:+175:1.1) arc (+90:+170:0.09587);
       \draw [green contour] (1.6,0) ++ (-10:0.09587) arc (-10:-90:0.09587) arc (-5:-175:1.1) arc (-90:-170:0.09587);

       \draw [green contour,-<] ([shift=(- 40:1.1)] 0.5,0) arc (- 40:- 45:1.1);
       \draw [green contour,-<] ([shift=(- 85:1.1)] 0.5,0) arc (- 85:- 90:1.1);
       \draw [green contour,-<] ([shift=(-130:1.1)] 0.5,0) arc (-130:-135:1.1);

       \draw [red contour,->] ([shift=(+140:1.1)] 0.5,0) arc (+140:+135:1.1);
       \draw [red contour,->] ([shift=(+ 95:1.1)] 0.5,0) arc (+ 95:+ 90:1.1);
       \draw [red contour,->] ([shift=(+ 50:1.1)] 0.5,0) arc (+ 50:+ 45:1.1);

       \node at (0,+0.78) [anchor=west] {$\scriptstyle +i$};
       \node at (0,-0.78) [anchor=west] {$\scriptstyle -i$};

       \draw [cross,blue] ([shift=(+120:1)] 0.5,0) +(+75:0.04) -- +(-105:0.04) +(+165:0.04) -- +(-15:0.04);
       \draw [cross,blue] ([shift=(-120:1)] 0.5,0) +(+195:0.04) -- +(+15:0.04) +(-75:0.04) -- +(+105:0.04);

       \draw [cross,violet] (-1.05,0) +(+135:0.04) -- +(-45:0.04) +(-135:0.04) -- +(+45:0.04);
       \draw [cross,violet] (-0.85,0) +(+135:0.04) -- +(-45:0.04) +(-135:0.04) -- +(+45:0.04);
       \draw [cross,violet] (-0.74,0) +(+135:0.04) -- +(-45:0.04) +(-135:0.04) -- +(+45:0.04);
       \draw [cross,violet] (-0.64,0) +(+135:0.04) -- +(-45:0.04) +(-135:0.04) -- +(+45:0.04);
       \draw [cross,violet] (-0.59,0) +(+135:0.04) -- +(-45:0.04) +(-135:0.04) -- +(+45:0.04);

       \draw [cross,violet] (+1.59,0) +(+135:0.04) -- +(-45:0.04) +(-135:0.04) -- +(+45:0.04);
       \draw [cross,violet] (+1.64,0) +(+135:0.04) -- +(-45:0.04) +(-135:0.04) -- +(+45:0.04);
       \draw [cross,violet] (+1.74,0) +(+135:0.04) -- +(-45:0.04) +(-135:0.04) -- +(+45:0.04);
       \draw [cross,violet] (+1.85,0) +(+135:0.04) -- +(-45:0.04) +(-135:0.04) -- +(+45:0.04);
       \draw [cross,violet] (+2.05,0) +(+135:0.04) -- +(-45:0.04) +(-135:0.04) -- +(+45:0.04);
       
       \fill (-0.515,0) circle (0.035);
       \fill (+1.52,0) circle (0.035);

       \fill (+0.5,0) circle (0.015);

       \draw [very thin,decorate,decoration={brace}] (0.5,0) -- (0,0);
       \draw [very thin,decorate,decoration={brace}] (1.5,0) -- (0.5,0);

       \node at (+0.25,0) [anchor=north] {$\scriptstyle \chi/\kappa$};
       \node at (+1,0) [anchor=north] {$\scriptstyle 1/\kappa$};

       \node at (-0.515,0.08) [anchor=west] {$\scriptstyle -1/s$};
       \node at (+1.52,0.08) [anchor=east] {$\scriptstyle +s$};

       \draw (1.6,1.1) node [anchor=center] {$x$} ++(-0.09,-0.09) ++ (0,0.18) -- ++(0,-0.18) -- ++(0.18,0);

     \end{tikzpicture}
   }

  \caption{For large $N$ we can use $\cot(\pi n) \approx \mp i$ to rewrite the integrals so that the contour splits into two parts, one in the upper and one in the lower half plane. Compared to the contours depicted in figure~\ref{fig:int-contours} the direction of integration is reversed in the upper (lower) half of the $n$ ($x$) plane. Following~\cite{Gromov:2007cd} we denote this integral by $\int_{-1/s}^{+s} = \frac{1}{2} \int_{\hat{\mathcal{D}}_+} + \frac{1}{2} \int_{\hat{\mathcal{D}}_-}$, where $\hat{\mathcal{D}}_{\pm}$ indicates the two halves of the contour shown in the figure.}
  \label{fig:int-contour-sgn}
\end{figure}
For large $N$, $\cot(\pi n)$ quickly approaches $\mp i$ in the upper/lower half plane. For the potential $\hat{\mathcal{V}}_k$ we then find
\begin{equation}
  \hat{\mathcal{V}}_k =
  + \frac{1}{2} \sum_{\hat{\imath}\hat{\jmath}} (-1)^F \int\limits_{-s^{-1}}^{+s} \frac{dy}{2\pi} ( \hat{p}_{\hat{\imath}}' - \hat{p}_{\hat{\jmath}}' ) \hat{\mathcal{V}}^{\hat{\imath}\hat{\jmath}}
  + \frac{1}{2} \sum_{\check{\imath}\check{\jmath}} (-1)^F \int\limits_{-s}^{+s^{-1}} \frac{dy}{2\pi} ( \check{p}_{\check{\imath}}' - \check{p}_{\check{\jmath}}' ) \hat{\mathcal{V}}^{\check{\imath}\check{\jmath}}
\end{equation}
where the integrals are to be taken in the upper and lower half planes along the contour in figure~\ref{fig:int-contour-sgn-x}.\footnote{%
  Note that this means we reverse the direction of the contour in the lower half plane. This sign change comes from the sign difference in the $\cot(\pi n)$ in the two half planes.%
}
Writing out the sums, the contribution from the first term in $\hat{\mathcal{V}}^{ij}$ completely cancels. For the potential $\hat{\mathcal{V}}_1^A $ we are left with
\begin{equation}
  \hat{\mathcal{V}}_1^A =
  + \frac{1}{2} \! \int\limits_{-s^{-1}}^{+s} \frac{dy}{2\pi} ( ( \hat{p}_2^A )' - ( \hat{p}_2^S )' ) \frac{\hat{\alpha}(x)}{x - y}
  + \frac{1}{2} \! \int\limits_{-s}^{+s^{-1}} \frac{dy}{2\pi} ( ( \check{p}_2^A )' - ( \check{p}_2^S )' ) \frac{\check{\alpha}(1/x)}{\frac{1}{x} - y}
\end{equation}
It is easy to check that the potentials added to the other hatted quasi-momenta are the same as the above up to a sign, as expected from the discussion in the beginning of this section.
To perform the integrals above we use the relations
\begin{equation}
  \begin{aligned}
    ( \hat{p}_2^A )' - ( \hat{p}_2^S )' &= + \partial ( G_{\hat{2}}(y) - G_{\check{2}}(1/y) ) , \qquad &
    G_{\hat{2}}(y) &= - \sum_{n=0}^{\infty} \hat{\mathcal{Q}}_{n+1} y^n ,
    \\
    ( \check{p}_2^A )' - ( \check{p}_2^S )' &= - \partial ( G_{\check{2}}(y) - G_{\hat{2}}(1/y) ) , \qquad &
    G_{\check{2}}(y) &= - \sum_{n=0}^{\infty} \check{\mathcal{Q}}_{n+1} y^n .
  \end{aligned}
\end{equation}
After factoring out a factor $\hat{\alpha}(x)$ we can expand the integrand at large $x$ and perform the integrals to obtain
\begin{equation}\label{eq:V-hat-result}
  \begin{split}
    \hat{\mathcal{V}}_1^A =
    &-\frac{\hat{\alpha}(x)}{2\pi} \sum_{m=1}^{\infty} \sum_{n=2}^{\infty} \frac{\hat{\mathcal{Q}}_n}{x^m} \Bigl(
    \frac{n-1}{m+n-2} - \frac{\delta_{m,1}}{2} \Bigr) \bigl( s^{m+n-2} - (-1/s)^{m+n-2} \bigr)
    \\
    &+\frac{\hat{\alpha}(x)}{2\pi} \sum_{m=1}^{\infty} \sum_{\substack{n=2 \\ n \neq m}}^{\infty} \frac{\check{\mathcal{Q}}_n}{x^m} \Bigl(
    \frac{n-1}{n-m} - \frac{\delta_{m,1}}{2} \Bigr) \bigl( (1/s)^{n-m} - (-s)^{n-m} \bigr)
    \\
    &-\frac{\hat{\alpha}(x)}{2\pi} \sum_{n=2}^{\infty} \frac{(n-1)\check{\mathcal{Q}}_n}{x^n} \log(s^2)
  \end{split}
\end{equation}
Note that the diagonal terms with $n=m$ are excluded in the summation in the second line. These terms are instead taking into account separately in the last line.

Setting $s=1$, the coefficients in the first line above agree with the result in equation (4.22) of~\cite{Beccaria:2012kb}. As discussed in that paper those coefficients are not anti-symmetric. This is in contradiction with unitarity of the dressing phase. In~\cite{Beccaria:2012kb,Beccaria:2012pm} it was argued that the symmetric terms appear due to the implicit choice of regulator in the algebraic curve approach. To compensate for the missmatch we will extract the part from the above result that is compatible with unitarity. From the first line above we obtain the correction to the phase $\sigma$, which we will denote by $\theta^{(1)}(x,y)$. Since unitarity requires this function to be anti-symmetric in its arguments we will write
\begin{equation}
  \theta^{(1)}(x,y) = \frac{1}{2} ( \vartheta^{(1)}(x,y) - \vartheta^{(1)}(y,x) ) ,
\end{equation}
where
\begin{equation}
  \!
  \vartheta^{(1)}(x,y) =
  -\frac{1}{2\pi} \sum_{m=1}^{\infty} \sum_{n=2}^{\infty} \frac{\hat{\alpha}(x) \hat{\alpha}(y)}{x^m y^n} \Bigl(
  \frac{n-1}{m+n-2} - \frac{\delta_{m,1}}{2} \Bigr) \bigl( s^{m+n-2} - (-1/s)^{m+n-2} \bigr) .
\end{equation}
Performing the anti-symmetrisation the phase $\theta^{(1)}(x,y)$ can then be written as
\begin{gather}
  \theta^{(1)}(x,y) = -\frac{1}{4\pi} \sum_{m=1}^{\infty} \sum_{n=m+1}^{\infty} c_{m,n} \Bigl( \frac{\hat{\alpha}(x) \hat{\alpha}(y)}{x^m y^n} - \frac{\hat{\alpha}(x) \hat{\alpha}(y)}{x^n y^m} \Bigr), \\
  c_{m,n} = \Bigr( \frac{n-m}{n+m-2} - \frac{\delta_{m,1} - \delta_{n,1}}{2} \Bigl) \bigl( s^{m+n-2} - (-1/s)^{m+n-2} \bigr) .
\end{gather}
For $s=1$ we recognise the coefficients in the phase of equation~(5.10) of~\cite{Borsato:2013hoa}. 

The one-loop correction to $\bar{\sigma}$ can be extracted from the second line of equation~\eqref{eq:V-hat-result}. We will denote this by $\bar{\theta}^{(1)}(\hat{x},\check{y})$, where the hat and check on the arguments are there to point out that this is the phase appearing in the S-matrix between one hatted and one checked excitation. In this case unitarity requires the phase to satisfy
\begin{equation}
  \bar{\theta}^{(1)}(\hat{x},\check{y}) + \bar{\theta}^{(1)}(\check{y},\hat{x}) = 0 .
\end{equation}
Since $\bar{\theta}$ describes scattering between two different excitations the above equation does not imply that this phase is anti-symmetric, because when exchanging the two particles we should also send $\chi \to -\chi$. Hence, we write
\begin{equation}
  \bar{\theta}(\hat{x},\check{y}) = \frac{1}{2} ( \bar{\vartheta}^{(1)}_{s} (\hat{x},\check{y}) - \bar{\vartheta}^{(1)}_{1/s} (\check{y},\hat{x}) ),
\end{equation}
where we have indicated the $s$-dependence.
From equation~\eqref{eq:V-hat-result} we find
\begin{equation}
  \begin{split}
    \bar{\vartheta}^{(1)}_{s} (\hat{x},\check{y}) = &
    +\frac{1}{2\pi} \sum_{m=1}^{\infty} \sum_{\substack{n=2 \\ n \neq m}}^{\infty} \frac{\hat{\alpha}(x) \check{\alpha}(y)}{x^m y^n} \Bigl(
    \frac{n-1}{n-m} - \frac{\delta_{m,1}}{2} \Bigr) \bigl( (1/s)^{n-m} - (-s)^{n-m} \bigr)
    \\
    &-\frac{1}{2\pi} \sum_{n=2}^{\infty} \frac{\hat{\alpha}(x) \check{\alpha}(y)}{x^n y^n} (n-1) \log(s^2) .
  \end{split}
\end{equation}
After imposing unitarity this leads to
\begin{gather}
  \bar{\theta}^{(1)}(\hat{x},\check{y}) = +\frac{1}{4\pi} \sum_{m=1}^{\infty} \sum_{n=1}^{\infty} 
  \bar{c}_{m,n} \frac{\hat{\alpha}(x) \check{\alpha}(y)}{x^m y^n},
  \\
  \bar{c}_{m,n} =
  \begin{cases}
    \displaystyle \Bigl( \frac{n+m-2}{n-m} - \frac{\delta_{m,1}-\delta_{n,1}}{2} \Bigr) \bigl( (1/s)^{n-m} - (-s)^{n-m} \bigr) & \text{for $n \neq m$} \\
    -2(n-1) \log s^2 & \text{for $n = m$} .
  \end{cases}
\end{gather}
Again setting $s=1$ we get back the coefficients in the pure RR case from equation~(5.10) of~\cite{Borsato:2013hoa}.

To derive another expression for the phases we directly perform the integrals in~\eqref{eq:V-hat-result} and impose unitarity to obtain
\begin{equation}\label{eq:result-dressing-phase}
  \begin{aligned}
    \theta^{(1)}(x,y) = - \frac{\hat{\alpha}(x) \hat{\alpha}(y)}{4\pi} \biggl[&
    \frac{1}{\kappa} \frac{(x+y)\bigl(1 - \frac{1}{xy}\bigr) - \frac{4\chi}{\kappa}}{(x-s)(x+s^{-1})(y-s)(y+s^{-1})} \frac{x+y}{x-y}
    \\ &\qquad
    + \frac{2}{(x-y)^2} \log\Bigl( \frac{y-s}{x-s} \frac{x+s^{-1}}{y+s^{-1}}\Bigr)
    \biggr] ,
    \\
    \bar{\theta}^{(1)}(x,y) = - \frac{\hat{\alpha}(x) \check{\alpha}(y)}{4\pi} \biggl[&
    \frac{1}{\kappa} \frac{(x-y)\bigl(1 + \frac{1}{xy}\bigr) - \frac{4\chi}{\kappa}}{(x-s)(x+s^{-1})(y+s)(y-s^{-1})} \frac{1+xy}{1-xy} 
    \\ &\qquad
    + \frac{2}{(1-xy)^2} \log\Bigl( \frac{x + s^{-1}}{x - s} \frac{y - s^{-1}}{y + s} s^2 \Bigr)
    \biggr] .
  \end{aligned}
\end{equation}
Using these expressions it is straight forward to check that
\begin{equation}
  \theta^{(1)}(x,y) + \bar{\theta}^{(1)}(x,1/y) = - \frac{i}{2} \frac{\hat{\alpha}(x) \hat{\alpha}(y)}{(x-y)^2} .
\end{equation}
The right hand side of the above equation perfectly matches the one-loop term in the crossing equation~\eqref{eq:crossing-equation-1}.
This serves as a non-trivial check that the expressions for the one-loop phase that we obtain are correct. Still it would be interesting to work out the phase in an independent way, for example from the scattering of giant magnons~\cite{Chen:2007vs,Abbott:2013ixa} where unitarity is manifest from the start.

The $\log$ terms of the dressing phases were derived in~\cite{Engelund:2013fja} using unitarity cut techniques.\footnote{%
  We thank Radu Roiban for pointing us to this reference.%
} %
By expressing the spectral parameters $\hat{x}$ and $\check{x}$ in terms of the BMN momenta $\hat{\mathrm{p}}$ and $\check{\mathrm{p}}$ using the relations
\begin{equation}
  \hat{\mathrm{p}} = \frac{\hat{x}}{(\hat{x}-s)(\hat{x}+s^{-1})} , \qquad
  \check{\mathrm{p}} = \frac{\check{x}}{(\check{x}+s)(\check{x}-s^{-1})} ,
\end{equation}
we find that the $\log$ terms in~\eqref{eq:result-dressing-phase} agree with the results of~\cite{Engelund:2013fja}.

\section{Classical circular string solutions}
\label{sec:circular-string}

Throughout this section we use the gauge $g_L\oplus g_R = g\oplus 1$, where $g\in \mathrm{psu}(1,1|2)$.
We parameterize $g$ following the notation of~\cite{Arutyunov:2003za} (used for the analog solution in $\AdS_5 \times \Sphere^5$).
\begin{equation}
  g = \begin{pmatrix}
    \sqrt{\frac{\mathcal{E}}{\varkappa}}e^{i \varkappa \tau} & \sqrt{\frac{\mathcal{S}}{w}}e^{i (w \tau+k \sigma)} & 0 & 0 \\
    \sqrt{\frac{\mathcal{S}}{w}}e^{-i (w \tau+k \sigma)} & \sqrt{\frac{\mathcal{E}}{\varkappa}}e^{-i \varkappa \tau} & 0 & 0 \\
    0 & 0 & \sqrt{\frac{\mathcal{J}_1}{\omega_1}}e^{i (\omega_1 \tau+m_1 \sigma)} & \sqrt{\frac{\mathcal{J}_2}{\omega_2}}e^{i (\omega_2 \tau+m_2 \sigma)} \\
    0 & 0 & \sqrt{\frac{\mathcal{J}_2}{\omega_2}}e^{-i (\omega_2 \tau+m_2 \sigma)} & \sqrt{\frac{\mathcal{J}_1}{\omega_1}}e^{-i (\omega_1 \tau+m_1 \sigma)} \\
  \end{pmatrix} .
\end{equation}
The parameters have to satisfy the following relations
\begin{equation}
  \begin{gathered}
    1 = \frac{\mathcal{E}}{\varkappa} - \frac{\mathcal{S}}{w} = \frac{\mathcal{J}_1}{\omega_1} + \frac{\mathcal{J}_2}{\omega_2}, \\
    k \mathcal{S} = m_1 \mathcal{J}_1 + m_2 \mathcal{J}_2, \\
    \omega_i^2 - m_i^2 +2 \chi \sum_{j=1}^{2} \omega_i (\sigma_1)^{ij} m_j = \nu^2, \\
    w^2 = \varkappa^2 + k^2 -2\varkappa k \chi, \\
    \frac{\mathcal{E}}{\varkappa } \varkappa ^2-\frac{\mathcal{S}}{w}\left(k^2+w^2\right) = \sum_{i=1}^2 \frac{\mathcal{J}_i}{\omega_i}\left(\omega_i^2+m_i^2\right).
  \end{gathered}
\end{equation}
By sending $\chi\to 0$, the equations reduce the known equations given in~\cite{Arutyunov:2003za}.

It is easy to find the quasi-momenta using the method given in~\cite{Dekel:2013dy,Dekel:2013kwa}.
However, let us consider some simple cases.
The simplest solution is the BMN solution where $\mathcal{S}=\mathcal{J}_2=m_1=0$ and $\omega_1=\mathcal{J}_1=\mathcal{E}=\varkappa$.
In this case the flat connection is constant (with respect to the worldsheet variables), and is given by
\begin{equation}
  A_{\sigma} (x) = \frac{i}{\kappa}\frac{x \mathcal{E}}{(x-s)(x-\frac{1}{s})}\mathrm{diag}(1,-1,1,-1)
  \oplus
  \frac{i}{\kappa}\frac{x \mathcal{E}}{(x+s)(x-\frac{1}{s})}\mathrm{diag}(-1,1,-1,1).
\end{equation}
Thus, the quasi-momenta is given by the eigenvalues of $-2\pi i A_\sigma$ which can be easily read from the above expression.
The only difference with the pure RR case where $s=1$ is the position of the poles and the residues which are rescaled by a factor equal to the pole position (\ie $s$ or $ 1/s$).
The quasi-momenta can be written as
\begin{equation}
  p_l(x) = (p(x),-p(x)|p(x),-p(x)||p(1/x),-p(1/x),p(1/x),-p(1/x))
\end{equation}
with
\begin{equation}
  p(x) = \frac{2\pi x \mathcal{E}}{\kappa(x-s)(x-\frac{1}{s})}.
\end{equation}

The next case we are going to consider is a one cut solution with the excitation on the sphere.
We take $m_1=-m_2=m$ and $\mathcal{J}_1=\mathcal{J}_2=\tilde{\mathcal{J}}/2$, $\varkappa=\mathcal{E}$, $\omega_2-\chi m_2=\omega_1-\chi m_1$ (we use the notation $\tilde{\mathcal{J}}$ rather than $\mathcal{J}$ in order to distinguish $\tilde{\mathcal{J}}$ from the angular momentum charge which we denote by $\mathcal{J}$).
At this point it is convenient to to introduce a new variable which can be interpreted as the average angular momentum on the sphere
\begin{equation}
  \Omega = \frac{1}{2}\left(\tilde{\mathcal{J}}+\sqrt{\tilde{\mathcal{J}}^2 + (2 m \chi)^2}\right),
\end{equation}
where in the pure RR case $\Omega \to \tilde{\mathcal{J}}$ is the angular momentum.
Using the new variable, the angular momenta are given by
\begin{equation}
  \omega_i = \Omega \pm m \chi,
\end{equation}
and the classical energy is given by
\begin{equation}
  \mathcal{E}^2 = m^2 \kappa^2 + \Omega^2.
\end{equation}
The flat connection matrix is given by four $2\times 2$ blocks, where it is enough to consider for the moment only the first two blocks.
The ``AdS'' block is identical to the BMN case described above.
The sphere's block is given by
\begin{equation}
  \hat A_\sigma^{S}(x) =
  \frac{i x }{ (x-s) \left(x+\frac{1}{s}\right)}
  \begin{pmatrix}
    m\left(\frac{\chi }{\kappa }- x\right) &  e^{2 i m (\sigma -\tau  \chi )} \frac{\sqrt{ \Omega ^2-(m \chi )^2}}{\kappa } \\
    e^{-2 i m (\sigma -\tau  \chi )} \frac{\sqrt{ \Omega ^2-(m \chi )^2}}{\kappa } & -m\left(\frac{\chi }{\kappa }- x\right)
  \end{pmatrix} .
\end{equation}
Next, we want to extract the quasi-momenta. However, the flat connection depends on $\sigma$ which might complicate things because of the path ordering integral.
Fortunately, we can use a simple gauge transformation in order to trivialize the integrations~\cite{Dekel:2013dy,Dekel:2013kwa}.
Notice that
\begin{equation}
S(\sigma)\hat A_\sigma^{S}(\tau,\sigma;x)S^{-1}(\sigma)
\end{equation}
is sigma independent if $S=e^{-i m \sigma \sigma_3}$. Thus, a Lax connection which yield an algebraic curve is given by
\begin{equation}
\hat L = \hat A_\sigma^{S}-\partial_\sigma S S^{-1}.
\end{equation}
Explicitly, the monodromy matrix is given by
\begin{equation}
\hat \Omega^{S}(x) = S^{-1}(2\pi) e^{\int_0^{2\pi} S(\sigma)\hat L S^{-1}(\sigma)} S(0) = e^{\int_0^{2\pi} S(\sigma)\hat L S^{-1}(\sigma)},
\end{equation}
where we used the fact that $S(2\pi)=S(0)=1$ since $m$ is an integer.
So finally, we only have to consider the eigenvalues of $2 \pi \hat L$.
Using $\partial_\sigma S S^{-1}=-i m \sigma_3$ the eigenvalues are given by
\begin{equation}
  \begin{aligned}
    \hat p_1^S(x) & = -\hat p_2^S(x) = \frac{2 \pi x K(1/x)}{\kappa (x-s) \left(x+\frac{1}{s}\right)} \\
    \check p_1^S(x) & = -\check p_2^S(x) = -\frac{2 \pi  x K(x)}{\kappa  (x+s)\left(x-\frac{1}{s}\right)}+2\pi m ,
  \end{aligned}
\end{equation}
where $K(x) = \sqrt{m^2 x \kappa  (x \kappa +2 \chi )+\Omega ^2}$.
The function $K(x)$ has branch points at $-\frac{\chi }{\kappa }\pm \frac{i}{m\kappa}\sqrt{\Omega ^2-(m \chi )^2}$. Hence, the branch cut is still parallel to the imaginary axis, but is shifted along the real axis by $-\chi/\kappa$. The length of the cut also gets bigger compared to the pure RR case where it is rescaled by $\sqrt{\frac{\Omega ^2-m^2 \chi ^2}{\mathcal{J}^2 \kappa ^2}}$.
The large $x$ expansion gives
\begin{equation}
  p_l(x)\simeq \frac{2\pi}{x \kappa}\left(\mathcal{E},-\mathcal{E}|\Omega,-\Omega||-\mathcal{E},\mathcal{E}|m \chi,-m \chi\right).
\end{equation}
Comparing with the general algebraic curve asymptotics in terms of the global charges we find $\mathcal{J} = \frac{1}{2}\left(\Omega+m\chi\right)$, $\mathcal{K} = \frac{1}{2}\left(\Omega-m\chi\right)$ so that $\mathcal{E}^2 = m^2 \kappa^2 + \left(\mathcal{J}+\mathcal{K}\right)^2$.

\section{The finite size giant magnon}
\label{sec:giant-magnon}

In this section we will calculate the leading finite size correction to the classical energy of the giant magnon~\cite{Hofman:2006xt} and its dyonic generalization~\cite{Chen:2006gea}. Finite size giant magnon solutions were first studied in $\AdS_5 \times \Sphere^5$~\cite{Arutyunov:2006gs}, and the corresponding energy correction has been computed using various methods~\cite{Arutyunov:2006gs,Astolfi:2007uz,Janik:2007wt,Hatsuda:2008gd,Minahan:2008re,Hatsuda:2008na,Ramadanovic:2008qd,Klose:2008rx}. The corresponding finite size solution in $\AdS_3 \times \Sphere^3 \times \Torus^4$ with a non-vanishing $B$-field was recently studied in~\cite{Ahn:2014tua}.

To construct giant magnons we consider the finite-gap equations in the $\algSU(2)$ sector. The quasi-momenta can then be written in terms of a single resolvent $G(x)$ and take the form
\begin{equation}
  \begin{aligned}
    - \hat{p}_2^A(x) = + \hat{p}_1^A(x) &= - \frac{1}{2} \frac{x \mathcal{D}}{(x - s)(x + s^{-1})} , \\
    - \hat{p}_2^S(x) = + \hat{p}_1^S(x) &= - \frac{1}{2} \frac{x \mathcal{D}}{(x - s)(x + s^{-1})} + G_{\text{mag}}(x), \\
    - \check{p}_2^S(x) = + \check{p}_1^S(x) &= + \frac{1}{2} \frac{x \mathcal{D}}{(x + s)(x - s^{-1})} + G_{\text{mag}}(1/x) - G_{\text{mag}}(0), \\
    - \check{p}_2^A(x) = + \check{p}_1^A(x) &= + \frac{1}{2} \frac{x \mathcal{D}}{(x + s)(x - s^{-1})} .
  \end{aligned}
\end{equation}
The giant magnon can be constructed from the resolvent~\cite{Minahan:2006bd,Vicedo:2007rp}
\begin{equation}
  G_{\text{mag}}(x) = +i \log \frac{x - X^+}{x - X^-} ,
\end{equation}
which describes a single logarithmic branch cut between the branch points $X^+$ and $X^-$, which we will take to be complex conjugates of each other. A single giant magnon is an unphysical string state carrying a world-sheet momentum, which is generically not an multiple of $2\pi$,
\begin{equation}
  p = -G_{\text{mag}}(0) = -i \log \frac{X^+}{X^-}.
\end{equation}
Note that $p$ is the world-sheet momentum of the magnon, which should not be confused with the quasi-moment $\hat{p}_i$ and $\check{p}_i$.
By expanding the quasi-momenta at large $x$ we find that the magnon carries Noether charges
\begin{equation}
  \begin{aligned}
    M = S - K &= -i\kappa h \bigl( X^+ + \frac{1}{X^+} - X^- - \frac{1}{X^-} \bigr) , \\
    E = D - J &= -i\kappa h \bigl( X^+ - \frac{1}{X^+} - X^- + \frac{1}{X^-} \bigr) .
  \end{aligned}
\end{equation}
Using the above expressions we find the dispersion relation
\begin{equation}
  D - J = \sqrt{ M^2 + 16 \kappa^2 h^2 \sin^2 \frac{p}{2} } .
\end{equation}
In~\cite{Hoare:2013lja} it was shown that in the mixed flux case the mass of the giant magnon takes the form
\begin{equation}\label{eq:gm-mass}
  M^2 = (\mathrm{Q} + 2\chi h p)^2 ,
\end{equation}
where $\mathrm{Q}$ is the bound state number of the dyonic magnon.

The derivation of the leading finite size corrections to the classical energy of the giant magnon in $\AdS_3 \times \Sphere^3 \times \Torus^4$ with mixed fluxes is almost identical to the corresponding calculation in $\AdS_5 \times \Sphere^5$. Hence, we keep the description here pretty brief and refer to the literature for the full details. To calculate the correction to the energy we need to deform the resolvent $G_{\text{mag}}(x)$. Such a deformation was found for the $\AdS_5 \times \Sphere^5$ giant magnon in~\cite{Minahan:2008re,Sax:2008in} and takes the form\footnote{%
  The same resolvent was applied to finite size giant magnons in $\AdS_4 \times \CP^3$ in~\cite{Lukowski:2008eq,Abbott:2009um}.%
}
\begin{equation}
  G_{\text{fin}}(x) = +2i \log \frac{ \sqrt{x-X^+} + \sqrt{x-Y^+} }{ \sqrt{x-X^-} + \sqrt{x-Y^-} } .
\end{equation}
Here we have introduced two short square root branch points between the points $X^\pm$ and $Y^\pm$, where the latter points are shifted a short distance $\delta \ll 1$ away from the original branch points
\begin{equation}
  Y^\pm = X^\pm ( 1 \pm i \delta e^{\pm i \phi} ) .
\end{equation}
When we add the extra square root branch cuts there is a back-reaction on the original branch points, which we take into account by making a further expansion
\begin{equation}
  X^\pm = X_{(0)}^\pm + X_{(1)}^\pm \delta + X_{(2)}^\pm \delta^2 .
\end{equation}
We parametrize the finite size magnon using the undeformed mass $M$ and the momentum $p$.
By requiring that $M$ and $p$ receive no corrections we find that the energy of the magnon up to quadratic order in $\delta$ takes the form
\begin{equation}
  D - J = \sqrt{M^2 + 16\kappa^2 h^2 \sin^2\frac{p}{2}} - \frac{\kappa^2 h^2 \sin^2\frac{p}{2}}{\sqrt{M^2 + 16\kappa^2 h^2 \sin^2\frac{p}{2}}} \cos(2\phi) \, \delta^2 .
\end{equation}
To express the parameters $\delta$ and $\phi$ in terms of the global charges we impose the finite-gap equation along the cut between $X^+$ and $Y^+$. In other word we solve the condition
\begin{equation}
  \hat{p}_1^S ( x + \epsilon ) - \hat{p}_2^S( x - \epsilon ) = 2\pi n ,
\end{equation}
where $x$ is a point on the cut and $\pm \epsilon$ implies that the quasi-momenta should be evaluated right above and below the cut.
From this equation we obtain
\begin{equation}
  \delta = 8 \sin\frac{p}{2} \exp\Bigl( - \frac{2 \sin^2\frac{p}{2} \sqrt{M^2 + 16\kappa^2 h^2 \sin^2\frac{p}{2}}}{16h^2 \sin^4\frac{p}{2} + ( M - 2\chi h \sin p)^2} D \Bigr)
\end{equation}
and
\begin{equation}
  \phi = -\frac{p}{2} + \pi n - \frac{1}{8\kappa h} \frac{M \cot\frac{p}{2} - 4 \chi h}{16h^2 \sin^4\frac{p}{2} + ( M - 2\chi h \sin p)^2} .
\end{equation}
Hence, the energy of the finite size magnon takes the form
\begin{multline}\label{eq:finite-size-magnon-energy}
  D - J = \sqrt{M^2 + 16\kappa^2 h^2 \sin^2\frac{p}{2}} \\
   -\frac{64 \kappa^2 h^2 \cos(2\phi) \sin^4\frac{p}{2}}{\sqrt{M^2 + 16\kappa^2 h^2 \sin^2\frac{p}{2}}} 
  \exp\Bigl( - \frac{2 \sin^2\frac{p}{2} \sqrt{M^2 + 16\kappa^2 h^2 \sin^2\frac{p}{2}}}{16h^2 \sin^4\frac{p}{2} + ( M - 2\chi h \sin p)^2} D \Bigr) .
\end{multline}
Note that the charge $D$ in the exponential on the right hand side should be taken is the $\AdS_3$ energy of the original magnon, and is a measure of the size of the magnon. In the strict giant magnon limit $D$ goes to infinity and the exponential factor vanishes. For large but finite $D$ the second term above gives an exponentially suppressed correction to the energy. This expression is very similar to the $\AdS_5 \times \Sphere^5$ result~\cite{Arutyunov:2006gs,Hatsuda:2008gd,Minahan:2008re}.
After identifying the mass $M$ as in~\eqref{eq:gm-mass} the above expression for the energy of the finite size magnon perfectly matches the result of~\cite{Ahn:2014tua}, up to some simple difference in notation.

The angle $\phi$ is a remnant of the missing momentum needed to construct a physical closed string state. As shown in~\cite{Minahan:2008re}, it is possible to construct physical string states consisting of multiple fundamental giant magnons. For such states the integer $n$ can be chosen in such a way that the $\phi$-dependent factor disappears. For more general multi particle solutions involving dyonic magnons $\phi$ is related to the relative angle between consecutive magnons.

\section{Conclusions}

We have constructed a set of finite-gap equations for string theory on $\AdS_3 \times \Sphere^3 \times \Torus^4$ with mixed RR and NSNS fluxes. 
Although the topological Wess-Zumino term, which carries the NSNS flux, breaks the $\Integers_4$ symmetry, we have used the $\Integers_4$ automorphism in an algebraic way such that the Lax connection satisfies standard $\Integers_4$ relations. This is an essential part of our derivation of the finite-gap equations. The resulting equations are similar to the case of pure RR flux but the positions of the poles in the source terms of the equations move away from $\pm 1$ for non-zero NSNS flux. We have also constructed a set of all-loop Bethe ansatz using the conjecture S-matrix of~\cite{Hoare:2013ida}. In the thermodynamic limit the Bethe equations reproduce the finite-gap equations derived from the world-sheet action. This serves as a check of the consistency between the conjectured S-matrix and the integrable world-sheet model.

Employing the same methods as in this paper it should be possible to construct finite-gap equations for other string theory backgrounds with mixed fluxes. String theory on $\AdS_3 \times \Sphere^3 \times \Sphere^3 \times \Sphere^1$ can also be supported by a combination of RR and NSNS fluxes. The corresponding action should be the same as the action considered in section~\ref{sec:super-coset-B-field-integrability} of this paper but with the superalgebra $\algD{\alpha}$ replacing $\algPSU(1,1|2)$. Some other classically integrable $\AdS_3$ and $\AdS_2$ backgrounds with mixed fluxes were also constructed in~\cite{Wulff:2014kja}.

The all-loop S-matrix for massive excitations was constructed based on the symmetry preserved by the ground state~\cite{Hoare:2013ida}. In the pure RR this algebra was recently derived directly from the gauge fixed world-sheet theory~\cite{Borsato:2014exa,Borsato:2014hja}. This derivation was essential for understanding how to obtain the full S-matrix including the massless modes. In the mixed flux case such a derivation could again be used for understanding how to include the massless modes in the integrability machinery. However, even for the massive sector such a derivation would be very interesting since it could help clarify the form of the dispersion relation of the world-sheet excitations, and in particular the origin of the momentum-dependent ``mass-term''.

The massive S-matrix further contains two undetermined scalar factors known as the dressing phases. In this paper we conjectured the form of the tree-level and one-loop part of these phases. To do this we imposed unitarity and crossing invariance, as well as the one-loop quantization of the algebraic curve. In the pure RR case the all-loop dressing phase was found in~\cite{Borsato:2013hoa} by solving the crossing equations. As an important part of that derivation the dispersion relation was written in a uniform way by introducing a rapidity torus (see also~\cite{Janik:2006dc} as well as the reviews~\cite{Arutyunov:2009ga,Vieira:2010kb} for the $\AdS_5 \times \Sphere^5$ case). In the mixed case such a derivation seems more complicated since the shortening condition satisfied by the spectral parameters $x^\pm$ is non-algebraic. It therefore seems that the rapidities naturally live on a Riemann surface with an infinite number of cuts rather than on a torus. It would be very interesting to understand the nature of this Riemann surface.

In our derivation of the one-loop dressing phases we manually impose unitarity. The origin of the regularisation ambiguity was discussed in~\cite{Beccaria:2012kb,Beccaria:2012pm}, but it would be good to further clarify how it can be resolved in the mixed flux case. It would also be interesting to confirm our results for the phases using a method where unitarity is mainfest from the start~\cite{Abbott:2013ixa}, and to confirm our results for the rational terms in the phase using perturbation theory~\cite{Sundin:2012gc,Rughoonauth:2012qd,Abbott:2012dd,Sundin:2013ypa,Abbott:2013ixa,Engelund:2013fja,Bianchi:2013nra,Sundin:2014sfa}.

In the parametrization we are using the Lax connection becomes degenerate in the pure NSNS limit. However, there are other choices of coefficients where the connection does not degenerate. It would be very interesting to study this limit more carefully and to understand how to construct the Lax connection directly in the pure NSNS theory.

\bigskip

\noindent\textbf{Note added:} After this paper was originally put on arXiv the article~\cite{Bianchi:2014rfa} was announced. Among other things, the authors calculate the dressing phases for the mixed flux string theory on $\AdS_3 \times \Sphere^3 \times \Torus^4$ using generalised unitarity cut techniques. The results they obtain confirm the form of the phases reported here.

\section*{Acknowledgments}

We would like to thank Ben Hoare, Joe Minahan, Bogdan Stefa\'nski, Per Sundin, Arkady Tseytlin, Konstantin Zarembo and Linus Wulff for interesting discussions. We also thank Arkady Tseytlin and Konstantin Zarembo for their comments on the manuscript.
O.O.S.'s  work was supported by the ERC Advanced grant No.\ 290456.\@ ``Gauge theory -- string theory duality''.

\appendix

\section{Details of the BMN string quantuzation}\label{app:BMNquantization}

In this appendix we give the details of the BMN string solution quantization using the algebraic curve following~\cite{Gromov:2007aq}.
The general procedure is explained in section~\ref{sec:semi-classical}, and the main result for the BMN solution quatization is given in~\ref{sec:BMN-quantization}.
We start be describing the quantization of fluctuations in the AdS sector, which is relatively simple and allows us to check the consistency of the procedure by  in the sense that the residues of the fluctuation poles are not imposed, but rather emerge by requiring consistency of the analytical structure and symmetries.
After that we turn to the quantization of the full spectrum of eight bosonic and fermionic fluctuations which yields to general energy fluctuations.

\subsection{AdS excitations}

For the AdS excitations we start with the following ansatz for the perturbed quasi-momenta
\begin{equation}
  \begin{aligned}\label{eq:adsQMpert}
    \delta \hat p^A_1(x) & = +\frac{s \delta\alpha}{x-s} - \frac{(1/s) \delta\alpha}{x+1/s}
    - \sum_n N^n_{\hat 1\hat 2}\frac{\hat \alpha(x^n_{\hat 1\hat 2})}{x-x^n_{\hat 1\hat 2}}
    + \sum_n N^n_{\check 1\check 2}\frac{\check \alpha(x^n_{\check 1\check 2})}{1/x-x^n_{\check 1\check 2}} + \hat a^A_1 ,
    \\
    \delta \hat p^S_1(x) & = \frac{s \delta\alpha_+}{x-s} - \frac{(1/s) \delta\alpha_+}{x+1/s},
  \end{aligned}
\end{equation}
and
\begin{equation}
  \begin{aligned}
    \delta\hat{p}^A_2(x) &= - \delta\hat{p}^A_1(x) , \qquad &
    \delta\check{p}^A_i(x) &= \delta\hat{p}^A_i(1/x) , \\
    \delta\hat{p}^S_2(x) &= - \delta\hat{p}^S_1(x) , \qquad &
    \delta\check{p}^S_i(x) &= \delta\hat{p}^S_i(1/x) .
  \end{aligned}
\end{equation}
The residues follow~\eqref{eq:reseq} and~\eqref{eq:poles}, while the constant should cancel the other constants in the large $x$ limit where we expect to find~\eqref{eq:QMasymp}.
The positions $x^n_{\hat{1}\hat{2}}$ and $x^n_{\check{1}\check{2}}$ are determined by the conditions~\eqref{eq:polepos}
\begin{equation}\label{eq:adsQMpertpolepos}
  \hat p_1^A(x^n_{\hat 1\hat 2}) - \hat p_2^A(x^n_{\hat 1\hat 2}) = 2\pi n \qquad
  \check p_2^A(x^n_{\check 1\check 2}) - \check p_1^A(x^n_{\check 1\check 2}) = 2\pi n.
\end{equation}
Notice that the overall minus sign in the last summation in~\eqref{eq:adsQMpert} comes from~\eqref{eq:reseq} and not from the inversion relation as in the $\AdS_5\times \Sphere^5$ case~\cite{Gromov:2007aq}.
(Since there are no sphere excitations we do not use the $\mathrm{AdS}$ superscripts.)
By requiring that the large $x$ expansion of the quasi-momenta starts at order $1/x$ we find the two conditions
\begin{equation}
  \hat{a}^A_1 = + \sum_n N^n_{\hat{1}\hat{2}} \frac{\hat{\alpha}(x^n_{\hat{1}\hat{2}})}{x^n_{\hat{1}\hat{2}}} , \qquad
  \hat{a}^A_1 = - \sum_n N^n_{\check{1}\check{2}} \frac{\check{\alpha}(x^n_{\check{1}\check{2}})}{x^n_{\check{1}\check{2}}} .
\end{equation}
This leads to the consistency condition
\begin{equation}\label{eq:bmn-ads-level-matching}
  \sum_n N^n_{\hat{1}\hat{2}} \frac{\hat{\alpha}(x^n_{\hat{1}\hat{2}})}{x^n_{\hat{1}\hat{2}}}
  + \sum_n N^n_{\check{1}\check{2}} \frac{\check{\alpha}(x^n_{\check{1}\check{2}})}{x^n_{\check{1}\check{2}}}
  =
  \frac{2\pi}{h \mathcal{J}} \Bigl( \sum_n N^n_{\hat{1}\hat{2}} n + \sum_n N^n_{\check 1\check 2} n \Bigr) = 0 ,
\end{equation}
which is just the level matching condition.

The next order in the large $x$ expansion leads to
\begin{equation}
\delta \alpha_\pm = 0 ,
\end{equation}
and to two expressions for $\delta\Delta$
\begin{equation}\label{eq:delta-Delta-expressions}
  \begin{aligned}
    \frac{1}{2} \delta\Delta &=
    \sum_n N^n_{\hat{1}{\hat{2}}} \bigl( \hat{\alpha}(x^n_{\hat{1}{\hat{2}}}) - 1 \bigr)
    + \sum_n N^n_{\check{1}{\check{2}}} \frac{\check{\alpha}(x^n_{\check{1}{\check{2}}})}{(x^n_{\check{1}{\check{2}}})^2} ,
    \\
    \frac{1}{2} \delta\Delta &=
    \sum_n N^n_{\check{1}{\check{2}}} \bigl( \check{\alpha}(x^n_{\check{1}{\check{2}}}) - 1 \bigr)
    + \sum_n N^n_{\hat{1}{\hat{2}}} \frac{\hat{\alpha}(x^n_{\hat{1}{\hat{2}}})}{(x^n_{\hat{1}{\hat{2}}})^2} ,
  \end{aligned}
\end{equation}
The sum of these expressions gives
\begin{align}
  \delta \Delta =
  \sum_n N^n_{\hat 1\hat 2} \left(\sqrt{\left(\frac{n}{\mathcal{J}}\right)^2+2 \chi \frac{n}{\mathcal{J}} + 1} - 1\right)
  + \sum_n N^n_{\check 1\check 2} \left(\sqrt{\left(\frac{n}{\mathcal{J}}\right)^2-2 \chi \frac{n}{\mathcal{J}} + 1} - 1\right),
\end{align}
For consistency the two expressions for $\delta\Delta$ in~\eqref{eq:delta-Delta-expressions} have to be equal. This gives the relation
\begin{equation}
  \sum_n N^n_{\hat{1}{\hat{2}}} \biggl( \hat{\alpha}(x^n_{\hat{1}{\hat{2}}}) \biggl( 1 - \frac{1}{(x^n_{\hat{1}\hat{2}})^2} \biggr) - 1 \biggr)
  -
  \sum_n N^n_{\check{1}{\check{2}}} \biggl( \check{\alpha}(x^n_{\check{1}{\check{2}}}) \biggl( 1 - \frac{1}{(x^n_{\check{1}\check{2}})^2} \biggr) - 1 \biggr)
  = 0 .
\end{equation}
It is enough for these sums to vanish when we impose level matching. This can be achieved by setting
\begin{equation}
  \hat{\alpha}(x^n_{\hat{1}{\hat{2}}}) = \frac{1 + C n}{1 - \frac{1}{(x^n_{\hat{1}\hat{2}})^2}} , \qquad
  \check{\alpha}(x^n_{\check{1}{\check{2}}}) = \frac{1 - C n}{1 - \frac{1}{(x^n_{\check{1}\check{2}})^2}} ,
\end{equation}
for some constant $C$. Using~\eqref{eq:adsQMpertpolepos} this gives us two expressions for the functions $\hat{\alpha}(x)$ and $\check{\alpha}(x)$.
If we further require these have only two poles in the $x$ plane we find that there are two possible values for the constant $C$. For $C=0$ both $\hat{\alpha}(x)$ and $\check{a}(x)$ are equal to $x^2/(x^2 - 1)$. This is the function used when quantizing the algebraic curve in $\AdS_5 \times \Sphere^5$~\cite{Gromov:2007aq}. If we instead set $C = \chi / \mathcal{J}$ we get back the expressions from~\eqref{eq:alphas}. Only in the second case is the level matching condition in~\eqref{eq:bmn-ads-level-matching} is satisfied.

\subsection{Full excitations spectrum}

Allowing all eight bosonic and fermionic excitations we start with the ansatz
\begin{equation}
  \begin{aligned}
    \delta \hat p^A_1(x) & = \frac{s \delta\alpha_+}{x-s} - \frac{(1/s) \delta\alpha_-}{x+1/s}
    - \sum_n (N^{AA}_{\hat 1\hat 2})^n\frac{\hat \alpha((x^{AA}_{\hat 1\hat 2})^n)}{x-(x^{AA}_{\hat 1\hat 2})^n}
    - \sum_n (N^{AS}_{\hat 1\hat 2})^n\frac{\hat \alpha((x^{AS}_{\hat 1\hat 2})^n)}{x-(x^{AS}_{\hat 1\hat 2})^n} \\
    &+\sum_n (N^{AA}_{\check1\check2})^n\frac{\check\alpha((x^{AA}_{\check1\check2})^n)}{1/x-(x^{AA}_{\check1\check2})^n}
    + \sum_n (N^{AS}_{\check 1\check 2})^n\frac{\check \alpha((x^{AS}_{\check 1\check 2})^n)}{1/x-(x^{AS}_{\check 1\check 2})^n}+ \hat a^A_1, \\
    \delta \hat p^A_2(x) & = \frac{s \delta\beta_+}{x-s} - \frac{(1/s) \delta\beta_-}{x+1/s}
    + \sum_n (N^{AA}_{\hat 1\hat 2})^n\frac{\hat \alpha((x^{AA}_{\hat 1\hat 2})^n)}{x-(x^{AA}_{\hat 1\hat 2})^n}
    + \sum_n (N^{SA}_{\hat 1\hat 2})^n\frac{\hat \alpha((x^{SA}_{\hat 1\hat 2})^n)}{x-(x^{SA}_{\hat 1\hat 2})^n} \\
    &-\sum_n (N^{AA}_{\check1\check2})^n\frac{\check\alpha((x^{AA}_{\check1\check2})^n)}{1/x-(x^{AA}_{\check1\check2})^n}
    - \sum_n (N^{SA}_{\check 1\check 2})^n\frac{\check \alpha((x^{SA}_{\check 1\check 2})^n)}{1/x-(x^{SA}_{\check 1\check 2})^n}+ \hat a^A_2, \\
    \delta \hat p^S_1(x) & = \frac{s \delta\alpha_+}{x-s} - \frac{(1/s) \delta\alpha_-}{x+1/s}
    + \sum_n (N^{SS}_{\hat 1\hat 2})^n\frac{\hat \alpha((x^{SS}_{\hat 1\hat 2})^n)}{x-(x^{SS}_{\hat 1\hat 2})^n}
    + \sum_n (N^{SA}_{\hat 1\hat 2})^n\frac{\hat \alpha((x^{SA}_{\hat 1\hat 2})^n)}{x-(x^{SA}_{\hat 1\hat 2})^n} \\
    &-\sum_n (N^{SS}_{\check1\check2})^n\frac{\check\alpha((x^{SS}_{\check1\check2})^n)}{1/x-(x^{SS}_{\check1\check2})^n}
    - \sum_n (N^{SA}_{\check 1\check 2})^n\frac{\check \alpha((x^{SA}_{\check 1\check 2})^n)}{1/x-(x^{SA}_{\check 1\check 2})^n}+ \hat a^S_1, \\
    \delta \hat p^S_2(x) & = \frac{s \delta\beta_+}{x-s} - \frac{(1/s) \delta\beta_-}{x+1/s}
    - \sum_n (N^{SS}_{\hat 1\hat 2})^n\frac{\hat \alpha((x^{SS}_{\hat 1\hat 2})^n)}{x-(x^{SS}_{\hat 1\hat 2})^n}
    - \sum_n (N^{AS}_{\hat 1\hat 2})^n\frac{\hat \alpha((x^{AS}_{\hat 1\hat 2})^n)}{x-(x^{AS}_{\hat 1\hat 2})^n} \\
    &+\sum_n (N^{SS}_{\check1\check2})^n\frac{\check\alpha((x^{SS}_{\check1\check2})^n)}{1/x-(x^{SS}_{\check1\check2})^n}
    + \sum_n (N^{AS}_{\check 1\check 2})^n\frac{\check \alpha((x^{AS}_{\check 1\check 2})^n)}{1/x-(x^{AS}_{\check 1\check 2})^n}+ \hat a^S_2.
  \end{aligned}
\end{equation}
The rest of the quasi momenta perturbations are given by the reflection relation, $\delta \check p(x) = \delta \hat p(1/x)$.
The position of the fermionic poles are found using the equations
\begin{equation}
  \begin{aligned}
    \hat p_1^A(x^n_{\hat 1\hat 2}) - \hat p_2^S(x^n_{\hat 1\hat 2}) & = 2\pi n \\
    \hat p_1^S(x^n_{\hat 1\hat 2}) - \hat p_2^A(x^n_{\hat 1\hat 2}) & = 2\pi n \\
    \check p_2^S(x^n_{\check 1\check 2}) - \check p_1^A(x^n_{\check 1\check 2}) & = 2\pi n \\
    \check p_2^A(x^n_{\check 1\check 2}) - \check p_1^S(x^n_{\check 1\check 2}) & = 2\pi n.
  \end{aligned}
\end{equation}
Using all the properties given above we find all the coefficients entering the quasi momenta
\begin{equation}
  \begin{aligned}
    \hat a^A_1 & = +\frac{2 \pi}{h \mathcal{J}}\sum_n((N^{AA}_{\check 1\check 2})^n+(N^{SA}_{\check 1\check 2})^n)n, \\
    \hat a^A_2 & = -\frac{2 \pi}{h \mathcal{J}}\sum_n((N^{AA}_{\check 1\check 2})^n+(N^{AS}_{\check 1\check 2})^n)n, \\
    \hat a^S_1 & = -\frac{2 \pi}{h \mathcal{J}}\sum_n((N^{SS}_{\check 1\check 2})^n+(N^{AS}_{\check 1\check 2})^n)n, \\
    \hat a^S_2 & = +\frac{2 \pi}{h \mathcal{J}}\sum_n((N^{SS}_{\check 1\check 2})^n+(N^{SA}_{\check 1\check 2})^n)n, 
  \end{aligned}
\end{equation}
\begin{equation}
  \begin{aligned}
    \delta \alpha_+ + \delta \alpha_- =&
    +\frac{2\pi}{h}\sum_n\bigg(
    (N^{SS}_{\check 1\check 2})^n
    +(N^{AS}_{\check 1\check 2})^n
    +(N^{SA}_{\hat 1\hat 2})^n
    +(N^{SS}_{\hat 1\hat 2})^n \\
    &-((N^{AS}_{\check 1\check 2})^n+(N^{SS}_{\check 1\check 2})^n)\sqrt{\left(\frac{n}{\mathcal{J}}\right)^2-2 \chi \frac{n}{\mathcal{J}} +1} \\
    &-((N^{SA}_{\hat 1\hat 2})^n+(N^{SS}_{\hat 1\hat 2})^n)\sqrt{\left(\frac{n}{\mathcal{J}}\right)^2+2 \chi \frac{n}{\mathcal{J}} +1}
    \bigg), \\
    \delta \alpha_+ - \delta \alpha_- =&
    +\frac{\pi}{h}\sum_n\bigg(
    (N^{AA}_{\check 1\check 2})^n
    -(N^{AS}_{\check 1\check 2})^n
    +(N^{SA}_{\check 1\check 2})^n
    -(N^{SS}_{\check 1\check 2})^n \\
    &+(N^{AS}_{\hat 1\hat 2})^n
    +(N^{AA}_{\hat 1\hat 2})^n
    -(N^{SA}_{\hat 1\hat 2})^n
    -(N^{SS}_{\hat 1\hat 2})^n
    \bigg), \\
    \delta \beta_+ + \delta \beta_- =&
    +\frac{\pi}{h}\sum_n\bigg(
    -(N^{AA}_{\check 1\check 2})^n
    -(N^{AS}_{\check 1\check 2})^n
    +(N^{SA}_{\check 1\check 2})^n
    +(N^{SS}_{\check 1\check 2})^n \\
    &-(N^{AS}_{\hat 1\hat 2})^n
    +(N^{AA}_{\hat 1\hat 2})^n
    -(N^{SA}_{\hat 1\hat 2})^n
    +(N^{SS}_{\hat 1\hat 2})^n
    \bigg), \\
    \delta \beta_+ - \delta \beta_- =&
    -\frac{2\pi}{h}\sum_n\bigg(
    (N^{AA}_{\check 1\check 2})^n
    +(N^{SA}_{\check 1\check 2})^n
    +(N^{AS}_{\hat 1\hat 2})^n
    +(N^{AA}_{\hat 1\hat 2})^n \\
    &-((N^{SA}_{\check 1\check 2})^n+(N^{SS}_{\check 1\check 2})^n)\sqrt{\left(\frac{n}{\mathcal{J}}\right)^2-2 \chi \frac{n}{\mathcal{J}} +1} \\
    &-((N^{AS}_{\hat 1\hat 2})^n+(N^{SS}_{\hat 1\hat 2})^n)\sqrt{\left(\frac{n}{\mathcal{J}}\right)^2+2 \chi \frac{n}{\mathcal{J}} +1}
    \bigg),
  \end{aligned}
\end{equation}
which finally yields
\begin{equation}
  \delta \Delta =
  \sum_{\text{all}~ij}\sum_n
  \left(
    \hat N^n_{ij} \left(
      \sqrt{\frac{n^2}{\mathcal{J}^2}+2 \chi \frac{n}{\mathcal{J}} +1} - 1
    \right)
    +
    \check N^n_{ij} \left(
      \sqrt{\frac{n^2}{\mathcal{J}^2}-2 \chi \frac{n}{\mathcal{J}} +1} - 1
    \right)\right).
\end{equation}

\bibliographystyle{nb}
\bibliography{refs,mixed-flux-finite-gap-paper}

\end{document}